\documentstyle[11pt,aaspp4,flushrt,psfig]{article}
\newcommand{\beq}{\begin{equation}}
\newcommand{\eeq}{\end{equation}}
\newcommand{\beqa}{\begin{eqnarray}}
\newcommand{\eeqa}{\end{eqnarray}}
\newcommand{\msun}{\hbox {$M_{\odot}$ }}
\newcommand{\dbv}{$\Delta ({\rm B} - {\rm V})$}
\newcommand{\dv}{\hbox {$\Delta \rm V^{TO}_{HB}$}}
\newcommand{\dvtwo}{\Delta {\rm V}}

\newcommand{\mvrr}{\hbox {$\rm M_v(RR)$}}
\newcommand{\mvto}{\hbox {$\rm M_v(TO)$}}
\newcommand{\mvbto}{\hbox {$\rm M_v(BTO)$}}
\newcommand{\ebv}{\hbox {$\rm E(B-V)$}}

\newcommand{\ea}{{\it et al.}}
\newcommand{\feh}{\hbox{$ [{\rm Fe}/{\rm H}]$}}
\newcommand{\kmsmpc}{\hbox{$ {\rm km}\, {\rm s}^{-1}\, {\rm Mpc}^{-1}$  
}}

\begin{document}
\rightline{{CERN-TH/97-121}}
\rightline{{CWRU-P4-97}}
\rightline{{astro-ph/9706128}}
\rightline{{June 1997 (revised July 97)}}

\title{\bf The Age Of Globular Clusters In Light Of
Hipparcos: Resolving the Age Problem?}

\author{Brian Chaboyer\altaffilmark{1,2}, P.~Demarque\altaffilmark{3}, 
Peter J.\ Kernan\altaffilmark{5}
and Lawrence M.\ Krauss\altaffilmark{4,5,6} }

\altaffiltext{1}{Steward Observatory, University of Arizona, Tucson,
AZ 85721~~E-Mail:
chaboyer@as.arizona.edu}

\altaffiltext{2}{Hubble Fellow}

\altaffiltext{3}{Department of Astronomy, and Center for Solar and  
Space
Research, Yale University, Box 208101, New Haven, CT 06520-8101
~E-Mail: demarque@astro.yale.edu}

\altaffiltext{4}{Theory Group, CERN, CH-1211 Geneva, Switzerland}

\altaffiltext{5}{Department of Physics,
Case Western Reserve University, 10900 Euclid Ave., Cleveland, OH
44106-7079~~E-Mail: pete@theory2.phys.cwru.edu;  
krauss@theory1.phys.cwru.edu}

\altaffiltext{6}{Also Department of Astronomy}

\begin{abstract}

We review five independent techniques which are used to set the
distance scale to globular clusters, including subdwarf
main sequence fitting utilizing the 
recent Hipparcos parallax catalogue.  
These data {\it together} all
indicate that globular clusters are farther away than previously
believed, implying a reduction in age estimates. We now adopt a best
fit value $\mvrr = 0.39 \pm 0.08 (stat)$ at $\feh =-1.9$ with an
additional uniform systematic uncertainty of $^{+0.13}_{-0.18}$.  This
new distance scale estimate is combined with a detailed numerical
Monte Carlo study (previously reported by Chaboyer
\ea\ 1996a)  designed to assess the uncertainty associated with the  
theoretical age-turnoff luminosity relationship in order to estimate
both the absolute age and uncertainty in age of the oldest globular
clusters.

Our best estimate for the mean age of the oldest globular clusters is
now $11.5\pm 1.3\,$Gyr, with a one-sided, 95\% confidence level lower
limit of 9.5 Gyr. This represents a systematic shift of over 2
$\sigma$ compared to our earlier estimate, due completely to the new
distance scale---which we emphasize is not just due to 
the Hipparcos data. This now provides a lower limit on the age of the
universe which is consistent with either an open universe, or a flat,
matter dominated universe (the latter requiring $H_0 \le
67\,\kmsmpc$).  Our new study also explicitly quantifies how
remaining uncertainties in the distance scale and stellar evolution
models translate into uncertainties in the derived globular cluster
ages. Simple formulae are provided which can be used to update our age
estimate as improved determinations for various quantities become
available.  Formulae are also provided which can be used to derive the
age and its uncertainty for a globular cluster, given the absolute
magnitude of the turn-off, or the point on the subgiant branch 0.05
mag redder than the turn-off.

\end{abstract}


\section{Introduction}
The absolute age of the oldest Galactic globular clusters (GCs)
currently provides the most stringent lower limit to the age of the
universe, and as such, provides a fundamental constraint on
cosmological models.  In particular, for some time the best GC age
estimates have been in direct contradiction with the maximum Hubble
age for the preferred cosmological model, a flat matter dominated
universe. The most recent comprehensive analyses suggested a lower
limit of approximately 12 Gyr for the oldest GC's in our galaxy (
e.g. Chaboyer, Demarque, Kernan \& Krauss 1996, hereafter
\cite{first}), which, for a flat matter dominated model, implies $H_0
\le 53\, \kmsmpc$, a value which is low compared to almost all
observational estimates.

Because of this apparent discrepancy, it remains critically
important to continue to re-evaluate the errors associated with the GC
age determination process itself.  GC age estimates are
obtained by comparing the results of theoretical stellar evolution
calculations to observed color magnitude diagrams.  The absolute
magnitude of the main-sequence turn-off (\mvto) has small
theoretical errors, and is the preferred method for obtaining the
absolute ages of GCs (e.g.\ \cite{renzini}).  Age
determination methods which utilize the color of the models, or post
main-sequence evolutionary models are subject to much larger
theoretical uncertainties, and do not lead to stringent age limits.

In recent years, a number of authors have examined the question of the
absolute age GCs (e.g.\ \cite{ckim}, \cite{mazz},
\cite{sal}) using different assumptions for the best available input
physics.  \cite{chaboyer} presented a table of absolute GC ages based
on a variety of assumptions for the input physics needed to construct
the theoretical age-\mvto\ relationship.  \cite{vand}\ have presented a
review of the absolute ages of the GCs, and by comparing results from
different authors, include an discussion on how various uncertainties
in the age dating process effect the final age estimate.

In order to obtain both a best estimate, and a well-defined lower
limit to the absolute age of the oldest GCs, we earlier adopted a
direct approach of running a Monte Carlo simulation.  In our Monte
Carlo, the various inputs into the stellar evolution codes were varied
within their inferred uncertainties, utilizing 1000 sets of
isochrones, and the construction of over 4 million stellar models.
From these theoretical isochrones, the age-\mvto\ relationship was
determined, and combined with an empirical calibration of
\mvrr\footnote{Throughout this paper, the terms \mvrr\ (V(RR)) and
${\rm M_V(HB)}$ (V(HB)) will be used interchangably, with the
understanding that ${\rm M_V(HB)}$ (V(HB)) refers to the mean
magnitude of the HB in the instability strip.}\ in order to calibrate
age as a function of the difference in magnitude between the main
sequence turn-off and horizontal branch (\dv).  This calibration was
used to derive the mean age of 17 old, metal-poor GCs using \dv.  The
principal result of this work was an estimate for the age of the
oldest GCs of $14.6\pm 1.7$ Gyr, with the one-sided 95$\%$ C.L.\ lower
bound of 12.1 Gyr (
\cite{first}) mentioned above.   Another important result was an  
explicit demonstration that the uncertainty in \mvrr\ overwhelmingly
dominated the uncertainty in the GC age determination.  We chose a
Gaussian distribution for the uncertainty in \mvrr\ because the data,
while scattered, appeared to be appropriately distributed about the
mean value, which we then determined to be $\mvrr =0.60$ at $\feh
=-1.9$, with an uncertainty of approximately $0.16$ at the $95\% $
confidence level.

Since this work was completed, the Hipparcos satellite has provided
improved parallaxes for a number of nearby subdwarfs (metal-poor
stars) (\cite{hipp}), the distance to a GC has been estimated using
white dwarf sequence fitting (\cite{renz2}), a number of new
astrometric distances to GCs have been published (\cite{rees}), and
improved theoretical horizontal branch models have become available
(\cite{hbmod}). This has lead us to critically re-evaluate the
globular cluster distance scale (and hence, the \mvrr\ calibration),
and update our estimate for the absolute age of the oldest GCs. 
We find, using the full Hipparcos catalogue along with the other
independent distance estimators
that all the data suggests that
this distance scale, and hence the GC age estimate
 have shifted by a significant amount, suggesting that the
dominant uncertainty in \mvrr\ was, and still is, not statistical but
rather systematic in character.

A detailed discussion of the globular cluster distance scale is
presented in \S \ref{sect2}.  The input parameters and distributions
in the Monte Carlo are presented in \S
\ref{sect3}.  The principal results of this paper are presented in \S
\ref{sect4}, which includes simple formulae which can be used to update  
the absolute age of the oldest GCs when improved estimates for the
various input parameters become available.  Finally, \S \ref{sect5}
contains a brief summary of our results, and a brief discussion of
their cosmological implications.

\section{The Globular Cluster Distance Scale\label{sect2}}
It is currently impossible to directly determine distances to GCs
using trigometric parallaxes. While such distance estimates may be  
available
in the future from micoarcsecond space astrometry missions  
(\cite{gaia},
\cite{sim}), at present a variety of secondary distance estimates are
all that is available for GC's.  The different techniques rely upon  
different
data and assumptions. As such, we have elected to review a number of
these techniques, and present a GC distance scale which is based on
combining 5 independent estimates.  To facilitate this,
we have reduced the various distance estimates to a calibration of
\mvrr.  This allows us to derive GC ages via the \dv\ method (see \S
\ref{sect4}).  As we are interested in absolute ages, we have focused  
our
attention on those techniques which rely upon the minimum number of
assumptions and thus hopefully should provide {\it a priori} the most  
reliable
absolute distances.

\subsection{Astrometric Distances}
A comparison of the proper motion and radial velocity dispersions
within a cluster allows for a direct determination of GC distances,
independent of reddening (\cite{cudworth}).  Although this method
requires that a dynamical model of a cluster be constructed, it is the
only method considered here which directly measures the distance to a
GC without the use of a `standard' candle.  The chief disadvantage of
this technique is its relatively low precision.  This problem is
avoided by averaging together the astrometric distances to a number
of different GCs.  Rees (1996) presents new astrometric distances to
eight GCs, along with two previous determinations.  As pointed out by
Rees, there are possibly large systematic errors in the dynamical  
modeling of M15, NGC 6397 and 47 Tuc. As such, these clusters will be excluded
in our analysis.  In addition Rees (private communication) cautions
that the distance to M2 will be revised soon to due to a new reduction  
of the M2 proper motions.  Excluding this cluster from the analysis
results in six clusters whose distances have been estimated
astrometrically.  Table \ref{tab1} tabulates the astrometric distances
from Rees (1996).  Unless otherwise noted, the numbers are those given
by Rees (1996)..  For the \feh\ values, we have given preference to the
high dispersion results of Kraft, Sneden and collaborators.   Taking
the weighted average of the \mvrr\ values listed in Table \ref{tab1}
results in $\mvrr = 0.59\pm 0.11$ at $<\feh> = -1.59$, where the
average \feh\ value has been calculated using the same weights as in
the \mvrr\ average.

\subsection{White Dwarf Sequence Fitting \label{white}}
Renzini \ea\ (1996) have utilized deep HST WFPC2 observations of NGC
6752 to obtain accurate photometry of the cluster white dwarfs.  They
have combined this with similar observations of local white dwarfs
with known parallaxes and masses (close to those in the cluster) to
derive the distance to NGC 6752 using a procedure similar to main
sequence fitting.  The derived distance modulus is ${\rm (m - M)_O} =
13.05\pm 0.10$ assuming $\ebv = 0.04$.  This reddening estimate is
from \cite{zinn}, and is identical to those found by \cite{burn} and
\cite{carney}.  NGC 6752 is a moderately metal-poor ($\feh = -1.54$
from \cite{zinnw}) cluster with $V(HB) = 13.63$ (tabulated by
\cite{cds}).  NGC 6752 has an extremely blue HB, thus, an estimate of
V(HB) relies upon an extrapolation of the observed photometry.  As
such the determination of V(HB) in NGC 6752 is rather uncertain, and
so have elected to take a rather generous error bar in the
determination of V(HB) of $\pm 0.1$ mag.  Combining the above
quantities yields $\mvrr = 0.45\pm 0.14$.

\subsection{Subdwarf Main Sequence Fitting}
Using parallaxes of nearby field stars, it is possible to define the
position of the ZAMS, and via a comparison to deep GC color magnitude
diagrams obtain a rather direct estimate of the distance to a cluster.
Unfortunately, the position of the ZAMS is a rather sensitive function
of metallicity, and there are few nearby subdwarfs.  Hence, there are
few metal-poor stars with well determined absolute magnitudes. 

The release of the Hipparcos data (\cite{hipp}) has improved this situation
somewhat, providing a large database of high quality parallax
measurements.  The Hipparcos catalogue contains over 100,000 stars, of
which nearly 21,000 stars have parallax errors less than 10\%.  The
Hipparcos catalogue has been searched for stars which are suitable for
GC main sequence fitting.  When selecting stars for use in main
sequence fitting, it is important to avoid potential biases due to
unresolved binaries and stars which are evolved off the ZAMS.  Known
or suspected binaries which are not resolved photometrically should be
avoided as both magnitudes and colors may be significantly altered by
the presence of a companion.  The use of stars which have evolved off
the ZAMS may lead to systematic errors in the derived distance
moduli, as it is not clear if GCs and metal-poor field stars are
exactly of the same age.  For example, a 2 Gyr age difference between a
calibrating subdwarf at ${\rm M_V}=5$ and a GC would lead to a
systematic error of 0.14 mag in the distance modulus (based on our
standard isochrones).  To be safe, we will only consider stars with
${\rm M_V} \ga 5.5$.  Fainter than this, the stars are essentially unevolved.

Many of the stars in the Hipparcos catalogue have large relative
parallax errors and are not useful for main sequence fitting.  For
these reason, we have elected to only consider stars with 
$\sigma_\pi/\pi < 0.10$. This stringent selection criterion was
selected to minimize potential Lutz-Kelker type biases 
(\cite{lutzk}, \cite{brown}).  The Hipparcos catalogue was searched
for stars which (a) have $\sigma_\pi/\pi < 0.10$, (b) are 
fainter than ${\rm M_V} \simeq 5.5$, and (c) are not known or suspected
Hipparcos binaries or variables.  This resulted in a list of 2618
stars, of which the great majority have near solar metallicity.  As we
are interested in the most metal-poor globular clusters, we require
stars with $\feh \la -1.0$.  To identify the metal-poor stars in the
Hipparcos sample, we have cross identified the above Hipparcos
subsample with a variety of \feh\ catalogues: the 1996 high resolution
spectroscopic catalogue \cite{cayrel}, the \cite{clla} catalogue, and
the measurements reported by \cite{gratton2} and \cite{pont}.  Stars
with $\feh \la -1.0$ were selected for further study. 
\cite{clla} and \cite{pont} obtained numerous radial velocity
measurements which could be used to identify potential binaries.
\cite{gratton2} have search for binaries based on an excess infrared
flux.  Any stars which were known or suspected binaries were removed
from our final list.  In total only 10 stars in the Hipparcos
catalogue pass our stringent selection criterion ($\sigma_\pi/\pi <
0.10$, ${\rm M_V} \ga 5.5$, $\feh \la -1.0$, and not known or
suspected binaries).  Given the small parallax errors in our sample,
Lutz \& Kelker (1973) type corrections are expected to be small
(\cite{brown}).  This issue is explored via a detailed Monte Carlo
analysis in Appendix \ref{appen1}, where it is concluded that the
sample is indeed free from systematic biases, and so no corrections
have been applied to the absolute magnitudes derived from the
Hipparcos parallax observations.  Another indication that possible
Lutz-Kelker type corrections are small for our sample is that the the
maximum $\sigma_\pi/\pi$ value is 0.08, well below our threshold of
0.10.  This makes it extremely unlikely that stars whose true parallax
are systematically smaller than the observed parallax are
preferentially included in our sample (the source of Lutz-Kelker type
biases).

Our subdwarf sample only has two stars in common with the sample of
\cite{pont}.  This is because the \cite{pont} sample only includes
five stars on the ZAMS (${\rm M_V} \ga 5.5$).  As discussed above,
even as small as a 2 Gyr age difference between evolved subdwarfs and
the GC will lead to systematic errors in the distance modulus of $\sim
0.14$ mag.  Of the unevolved  stars in the \cite{pont} sample, three
are known binaries, which we do not use.  \cite{pont} apply an average
binary correction of $+0.375$ mag to the 6 binaries in their total
sample.  The Poisson (root $N$) noise in this correction is $\pm 0.15$
mag.  If one has a large sample of binaries ($\ga 30$), then the
approach taken by \cite{pont} to include average binary corrections is
sound.  However, given that the small number statistics in the present
sample results in a very large error in the binary correction, we
believe it is best not to use the binaries.

Theoretical models predict that the location of the ZAMS is a
sensitive function of metallicity.  Even with the Hipparcos data, the
current observations are not accurate (or numerous) enough to
empirically derive the ZAMS location as a function of metallicity.
There are only four subdwarfs whose absolute magnitudes are known to
within $\pm 0.1$ mag.  Unfortunately, the colors predicted by the
models are still rather uncertain, and so we do not have a reliable
calibration of how the location of the ZAMS changes as a function of
metallicity.  Thus one should ensure that the mean metallicity of  
subdwarf sample used in the main sequence fitting should be as close as
possible to the metallicity of the GC.  This requires accurate
metallicity determinations.  

We have searched the literature for abundance analyses, based upon
high dispersion, high signal to noise spectrum.  King (1997) has
performed a detailed abundance analysis of HD 134439 and HD 134440 (a
common proper motion pair).  Rather surprisingly, King finds that the
abundances of the $\alpha$-capture elements are consistently some
$\sim 0.3$ dex below the vast majority of metal-poor field stars, and
those observed in GC giant stars. Due to their relatively high
abundance, theoretical models predict that the $\alpha$-capture
elements play an important role in determining the position of a star
in the color magnitude diagram.  Given the peculiar abundances in
these two stars, we have elected not to use them in main sequence
fitting.  The calibrating subdwarf data for the remaining 8 stars is
presented in Table \ref{tab2}.  Most of the data in Table
\ref{tab2} has been taken from the Hipparcos catalogue.  The
reddening estimates are from \cite{clla} when available, or
\cite{pont}.  The \feh\ values are discussed in detail below.  

The \feh\ abundance of HD 193901 has been determined by a number of
groups.  Recent values of \feh\ are $-1.00$ (\cite{gratton}), $-0.98$
(\cite{axer}) and $-1.22$ (\cite{tomkin}).  The higher abundances derived
by \cite{axer} and \cite{gratton2} may be largely due to the different
effective temperature scales adopted by these authors compared to
\cite{tomkin}.  This is still a matter of active debate, so we have
elected to simple average the above \feh\ values.  In our main
sequence fitting analysis (\S \ref{sectn6752} and \S \ref{sectm5}),
we explore the consequences of the various \feh\ values. HD 145417 has
not been extensively studied, and the only spectroscopic metallicity
determination is $\feh = -1.15\pm 0.13$ (\cite{gratton2}).

Balachandran \& Carney (1996) have presented a detailed abundance
analysis of HD 103095 (Groombridge 1830), the subdwarf with 
the best determined absolute magnitude.  They found $\feh = -1.22\pm
0.04$.  Other abundances appeared to be typical of metal-poor stars.  
This is very similar to the value obtained by \cite{gratton2}
($\feh = -1.24\pm 0.07$).  The Balachandran \& Carney (1996) value is
adopted in this work. The abundance of HD 120559 has been determined
to be $\feh = -1.23\pm 0.07$ (\cite{axer}).  \cite{tomkin} have
found $\feh = -1.45$ for HD 126681.  

The metallicity of BD+59 2407 is $\feh = -1.60\pm 0.16$
(\cite{gratton2}) based on high signal to nose data. 
\cite{clla} found $\feh = -1.91$ in their low signal to noise
data.  An examination of the \cite{gratton2} and \cite{clla}
abundances indicates that the former are systematically more
metal-rich than the later.  Gratton \ea\ (1997) found that on average,
their \feh\ values were +0.34 dex more metal rich than \cite{tomkin}.
Once again, part of this difference is attributable to the different
effective temperature scales.  We have again 
elected to adopt the average
value $\feh = -1.75$, and will discuss the different \feh\ scales in
\S \ref{sectn6752} and \S \ref{sectm5}.

A detailed abundance analysis of HD 25329 has been presented by
Beveridge \& Sneden (1994).  They found $\feh = -1.84\pm 0.05$, while
\cite{gratton2} report $\feh = -1.69\pm 0.07$. Beveridge \& Sneden
(1994) note that HD 25329 is a N-enhanced star; only $\sim 3\%$ of
observed halo dwarfs are N-enhanced.  However, they find that the
relative abundances of $\alpha$-capture and iron peak elements are
normal for metal-poor stars.  Nitrogen comprises only $\sim 3\%$ of
the mass fraction of the heavy elements in a star.  Thus, the fact
that HD 25329 is N-enhanced is unlikely to affect its position on the
color magnitude diagram, and so it will be used in the main sequence
fitting.  Taking a simple average of the above two \feh\ determinations
results in $\feh = -1.76$.

\begin{figure}
\centerline{\psfig{figure=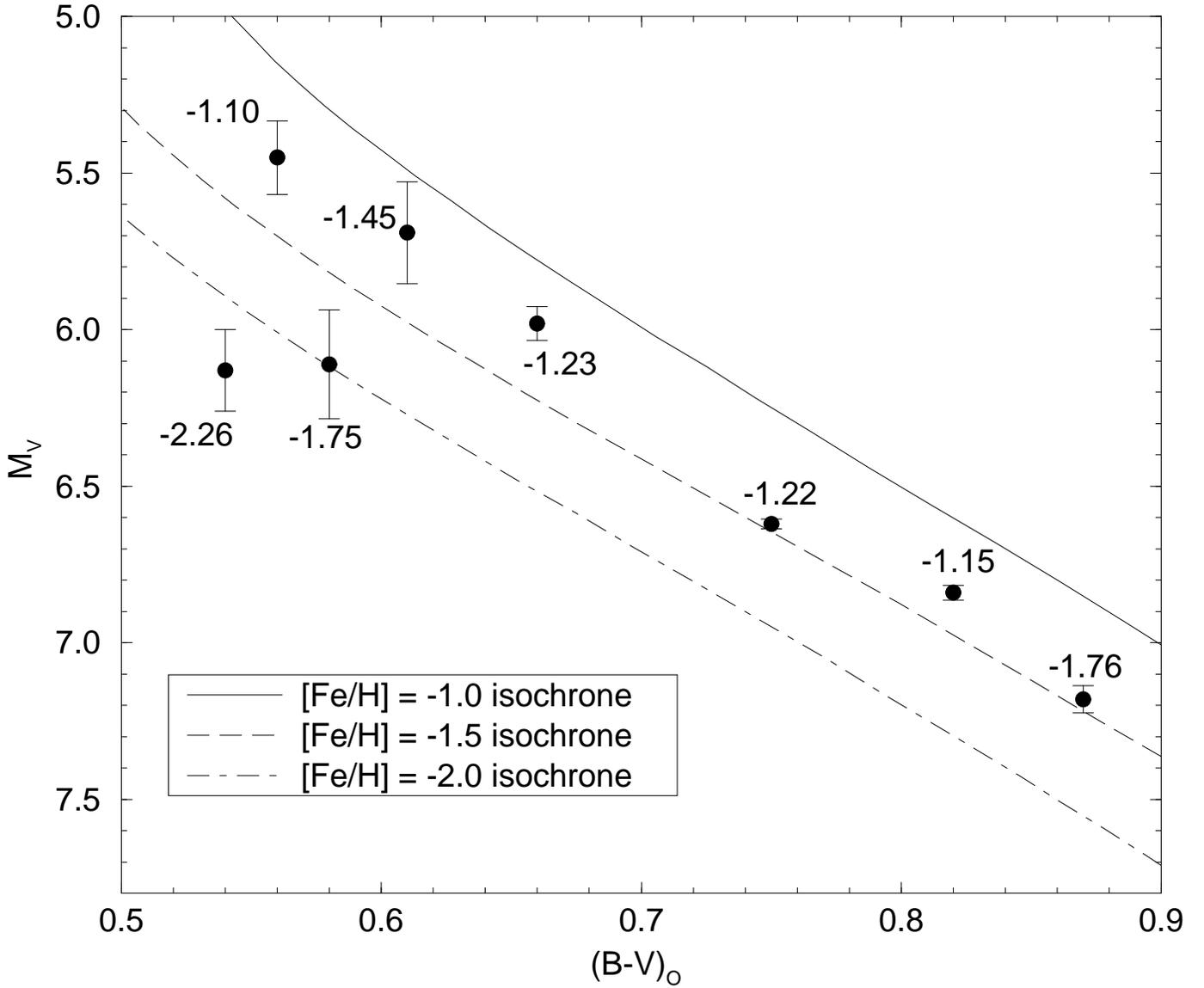,height=18.0cm,angle=270}  }
\caption{Unevolved ($M_V \ga 5.5$), metal-poor ($\feh \la -1.0)$ stars
in the Hipparcos catalogue which have very good parallaxes
($\sigma_{\pi}/\pi < 0.10$) and which are not known binaries are
compared to our isochrones.  Each Hipparcos star (points with error
bars) is labeled with its spectroscopic  \feh\ value.}
\label{fighr}
\end{figure}

Finally,  CPD-80 349 has $\feh = -2.26\pm 0.2$ (\cite{pont}).  This is
based on low signal to noise spectra, and is on the same system as
\cite{clla}. This is apparently the most metal-poor star in the
Hipparcos catalogue (see Figure \ref{fighr}), and an improved
abundance determination would be of great benefit. This star has 
$\ebv = 0.02$ (\cite{pont}).  Given the poor quality of the abundance
determination, and the fact that it is difficult to determine the
reddening for a single star, we have elected not to use CPD-80 349 in
our main sequence fits.  

Figure \ref{fighr} presents the HR diagram for the calibrating
subdwarf data, along with a comparison with our standard isochrones.
Most of the stars have metallicities in the range $-1.1$ to $-1.5$ and
provide a nice calibration of the ZAMS in this \feh\ range.  The
position of HD 25329 ($\feh = -1.76$) is somewhat surprising, as it
lies along the same isochrone as HD 103095 ($\feh = -1.22$).  Both of
these stars have very well determined metallicities and parallaxes.
Note that BD+59 2407 ($\feh = -1.75$) does not lie along the same
isochrone as HD 25329.  This could be due to an error in the
reddening or parallax of BD+59 2407.  Alternatively, it suggests that
HD 25329 is anomalously bright for its metallicity and color.
Clearly more data is needed to differentiate between these
hypothesis.  Unfortunately, an inspection of the Hipparcos catalogue
reveals there are no candidate metal-poor, unevolved  single stars
with $\sigma_\pi/\pi < 0.10$ which are likely to have $\feh < -1.5$.

Given the \feh\ values of the calibrating subdwarfs, accurate GC distances
using main sequence fitting can be obtained for GC with $-1.8 \la \feh
\la -1.1$.  To perform main sequence fitting, accurate photometry well
below the main sequence turn-off is required.  In addition, the
cluster \feh\ value must be well determined.  Given these
restrictions, accurate main sequence fitting distances can only be
determined to three globular clusters M5, M13 and NGC 6752.  NGC 288
and 362 are not included, as moderate resolution spectra of a few
cluster giants yield \feh\ values which are considerably different
(\cite{gratton}) from the Zinn \& West (1984) values.  Higher
dispersion spectra of a number of stars in each cluster are required
to determine accurate \feh\ abundances for these two clusters.  Our
approach to subdwarf fitting differs significantly from that adopted
by Reid (1997), Gratton \ea\ (1997) and Pont \ea\ (1997) in that (a)
we do not use binary, evolved or chemically peculiar subdwarfs and (2)
we do not make theoretical `color' corrections to the subdwarf data to
account for metallicity differences between the GC and subdwarfs, but
ensure that the mean metallicity of the subdwarf sample is nearly
identical to the GC.  The subdwarf fitting results for each of the
three clusters (NGC 6752, M5, M13) are discussed in turn.

\subsubsection{NGC 6752\label{sectn6752}}
High resolution spectra of three giants yields $\feh = -1.58$
(\cite{minniti}), while the six giants studied by \cite{ndc} yield
$\feh = -1.52$. \cite{gratton} obtained data for 4 other
giants, and re-analyzed the above data to obtain $\feh = -1.42$.
Averaging these three abundance determinations, we adopt $\feh =
-1.51$. The reddening is $\ebv = 0.04$, as discussed in \S
\ref{white}.  Subdwarfs with $-1.23 \le \feh \le -1.76$ were used in
the weighted, least squares fit. The mean abundance of these subdwarfs
(using the same weighted as in the least squares fit to the NGC 6752
fiducial) is $\feh = -1.55$, very similar to our adopted \feh\
abundance of NGC 6752.  

The distance to NGC 6752 was determined using a weighted least squares
fit to the deep photometry of this cluster as presented by
\cite{penny}.  The weights for the fit were the errors in the absolute
magnitudes of the subdwarfs ($\sigma_{M_V}$, presented in Table
\ref{tab2}).  These absolute magnitude errors only include the
parallax errors.  To allow for errors in the photometry, an error of
$\pm 0.02\,$mag was added in quadrature with the $\sigma_{M_V}$
tabulated in Table \ref{tab2} when performing the fit.   The
resultant distance modulus is ${\rm (m-M)_V}= 13.33\pm 0.04$ mag, where
this error represents the error associated with the weighted least
squares fit of the NGC 6752 fiducial to the subdwarf data.  To this
error, one must add in errors associated with the reddening, and allow
for possible metallicity errors.  An uncertainty in the reddening of
$\pm 0.01$ translates into an error in the derived distance modulus of
$\pm 0.05$.

Due to the possible systematic uncertainties in the metallicity
abundances of the subdwarfs and NGC 6752 we have examined various
possibilities, in order to determine how a possible mis-match
between the metallicity of NGC 6752 and the calibrating subdwarfs
might affect the distance modulus estimates.
\begin{enumerate}
\item Adopting the \cite{gratton} abundance for NGC 6752 ($\feh =
-1.42$), and using subdwarfs with $-1.22 \le \feh \le -1.76$.  The
weight of HD 103095 in the fit was decreased by increasing its 
$\sigma_{M_V}$ error to $\pm 0.045$, ensuring that the mean
weighted mean metallicity of the 5 calibrating subdwarfs was $\feh =
-1.42$.  The resultant distance modulus is 
${\rm (m-M)_V}= 13.31\pm 0.03$ mag.

\item Adopting the \cite{gratton} abundance for NGC 6752 along with
the \cite{gratton2} and \cite{axer} abundances for the subdwarfs.
This was done as the \cite{axer} and \cite{gratton2} subdwarf
abundances are systematically more metal-rich than other
determinations.  HD 126681 does not have an abundance
determinations by \cite{gratton2} or \cite{axer} and was removed from
the list.  The 3 stars with $-1.24 \le \feh \le -1.69$ were used in
the fit (HD 103095, 25329 and BD+59 2407).  Once again, the 
weight of HD 103095 in the fit was decreased by increasing its 
$\sigma_{M_V}$ error to $\pm 0.038$, ensuring that the mean
weighted mean metallicity of the calibrating subdwarfs was $\feh =
-1.42$.  The resultant distance modulus is ${\rm (m-M)_V}= 13.25\pm
0.03$ mag.

\item Assuming that a systematic zero-point error ($+0.20$ dex) exists 
between the subdwarf \feh\ determinations and NGC 6752 implying that NGC
6752 has a metallicity of $\feh =  -1.71$ (in the subdwarf \feh\
system).  This results in the use of 3 stars with $-1.45 \le \feh \le
-1.76$, and decreasing the weight of HD 25329 in the fit by increasing its
$\sigma_{M_V}$ error to $\pm 0.073$\,mag.  The derived distance
modulus is ${\rm (m-M)_V}= 13.24\pm 0.06$ mag.

\item Assuming that a systematic zero-point error ($-0.20$ dex) exists 
between the subdwarf \feh\ determinations and NGC 6752 implying that NGC
6752 has a metallicity of $\feh =  -1.31$ (in the subdwarf \feh\
system). The three subdwarfs with $-1.22 \le \feh \le -1.45$ with
equal weighting (implying a mean $\feh = -1.30$) were used in the fit.  
The resultant distance modulus is ${\rm (m-M)_V}= 13.42\pm 0.09$ mag.

\item Assuming the anomalous position of HD 25329 in the Figure
\ref{fighr} is due to an incorrect abundance determination, and so
removing HD 25329 from the fit. The three subdwarfs with $-1.23 \le
\feh \le -1.75$ were used in the fit, and the $\sigma_{M_V}$ error in
HD 120559 was increased to $\pm 0.22$ so that the weighted mean
subdwarf \feh\ was $-1.51$.  The derived distance modulus is ${\rm
(m-M)_V}= 13.30\pm 0.11$ mag.

\item Assuming that the \feh\ determination of BD+59 2407 is in error,
and removing it from the fit.  This results in the use of the three
subdwarfs with $-1.23 \le \feh \le -1.76$ and a distance modulus of 
${\rm (m-M)_V}= 13.34\pm 0.04$ mag.

\end{enumerate}

The maximum change in the derived distance modulus is $\pm 0.09$ mag,
which we take to be the 1-$\sigma$ error in the distance modulus due
to possible metallicity errors.  Adding the metallicity, reddening and
fitting errors together in quadrature yields a distance modulus of
${\rm (m-M)_V}= 13.33\pm 0.11$.  As discussed in \S \ref{white}, ${\rm
V(HB)} = 13.63\pm 0.1$, and so $\mvrr = 0.30\pm 0.15$ from the
subdwarf distance modulus.  This subdwarf visual distance modulus
corresponds to ${\rm (m-M)_O} = 13.20\pm 0.11$ (with ${\rm A_V} =
3.2$), which is within 1-$\sigma$ of the distance obtained from the
white dwarfs ${\rm (m - M)_O} =13.05\pm 0.10$ (\cite{renz2}).

\subsubsection{M5 \label{sectm5}}
High dispersion spectroscopic analysis indicates that this cluster has
$\feh = -1.17$ (\cite{snedenm5}).  The reddening is $\ebv = 0.03$, as
summarized by \cite{reid}.  A deep color magnitude diagram for this
cluster has been presented by \cite{sand}.  Subdwarfs with $-1.07 \le
\feh \le -1.23$ were used in the weighted, least squares fit. The
weighted mean abundance of these subdwarfs $\feh = -1.19$, very
similar to our adopted \feh\ abundance of NGC 6752. Subdwarf fitting
yields a distance modulus of ${\rm (m-M)_V} = 14.51 \pm 0.02$.  To
this error, one must add in errors associated with the reddening, and
allow for possible metallicity errors.  An uncertainty in the
reddening of $\pm 0.01$ translates into an error in the derived
distance modulus of $\pm 0.05$.

Due to the possible systematic uncertainties in the metallicity
abundances of the subdwarfs and M5, we have once again examined 
examined the effects that various scenarios for metallicity errors
have on the derived distance modulus.  
\begin{enumerate}
\item Adopting the  \cite{gratton2} and \cite{axer} abundances for the
subdwarfs, and using the 4 subdwarfs with $-0.99 \le \feh \le -1.23$
(HD 103095, 120559, 145417 and 193901) resulting in a mean metallicity of the
subdwarfs of $\feh = -1.20$ and ${\rm (m-M)_V} = 14.51 \pm 0.03$.

\item Assuming that a systematic zero-point error (+0.20 dex) exists 
between the subdwarf \feh\ determinations and M5 implying that M5 has
a metallicity of $\feh = -1.37$ (in the subdwarf \feh\ system).  This
results in the use of 5 stars with $-1.22 \le \feh \le -1.76$, and
decreasing the weight of HD 103095 in the fit by increasing its
$\sigma_{M_V}$ error to $\pm 0.032$\,mag.  The derived distance
modulus is ${\rm (m-M)_V}= 14.49\pm 0.02$ mag.

\item Assuming that the $\feh$ value for HD 103095 is in error, and so
removing it from the fit.  This results in the use of three subdwarfs 
($-1.07 \le \feh \le -1.23$ and ${\rm (m-M)_V}= 14.58\pm 0.03$ mag.

\item Removing HD 193901 from the fit (leaving HD 103095, 120559 and
145417 with a mean metallicity of $\feh = -1.20$).  The resultant
distance modulus is ${\rm (m-M)_V}= 14.51\pm 0.02$ mag.

\item Removing HD 145417 from the fit, and giving equal weight to the
remaining three stars (HD 103095, 120559 and 193901) to ensure a mean
$\feh = -1.17$.  The derived distance modulus is ${\rm (m-M)_V}=
14.58\pm 0.07$ mag.

\item Only using HD 145417 ($\feh = -1.15$) in the fit, resulting in 
 ${\rm (m-M)_V}= 14.56\pm 0.03$ mag.

\end{enumerate}

The maximum change in the derived distance modulus is $\pm 0.07$ mag,
which we take to be the 1-$\sigma$ error in the distance modulus due
to possible metallicity errors.  Adding the metallicity, reddening and
fitting errors together in quadrature yields a distance modulus of
${\rm (m-M)_V}= 14.51\pm 0.09$.   Utilizing ${\rm V(RR)}
= 15.05\pm 0.02$ (\cite{reid2}), results in $\mvrr = 0.54\pm 0.09$.

\subsubsection{M13 \label{sectm13}}
High dispersion spectroscopic analysis indicates that this cluster has
$\feh = -1.58$ (\cite{kraft}).  The reddening is $\ebv = 0.02$
(\cite{zinnw}). A deep color magnitude diagram and fiducial has been
obtained by \cite{richer}.  This is a somewhat difficult metallicity
to deal with, as none of the calibrating subdwarfs has a metallicity
near $\feh = -1.58$.  We have explored a number of possible options
for a subdwarf sample selection.
\begin{enumerate}
\item Utilizing the subdwarfs with $ -1.23\le \feh \le
-1.76$, results in a weighted mean $\feh = -1.55$ for the subdwarfs
and ${\rm (m-M)_V} = 14.54\pm 0.04$ is obtained from a weighted fit to
the fiducial.  

\item The subdwarf \feh\ range is restricted to 
$ -1.45\le \feh \le -1.76$ and the stars are equally weighted
(resulting in a mean $\feh = -1.65$) then the derived distance modulus
is ${\rm (m-M)_V} = 14.46\pm 0.10$

\item A systematic offset error of $-0.2$ dex is assumed between the
subdwarfs and M13, implying that M13 has $\feh = -1.78$ in the
subdwarf system.  In this case, only the two  stars with 
$\feh = -1.75$ and $-1.76$ are used, resulting in 
${\rm (m-M)_V} = 14.39\pm 0.05$

\item A systematic offset error of $+0.2$ dex is assumed between the
subdwarfs and M13, implying that M13 has $\feh = -1.38$ in the
subdwarf system. In this case subdwarfs with 
$ -1.22\le \feh \le -1.76$ are used, and the weight of HD 103095 in
the fit is decreased by increasing its $\sigma_{M_V}$ error to $\pm
0.034$ (to ensure a subdwarf mean $\feh = -1.38$).  The derived
distance modulus is ${\rm (m-M)_V} = 14.51\pm 0.02$
\end{enumerate}

The derived distance moduli vary from 14.39 -- 14.54.  We have elected
to adopt the mid-point as our best value, and utilize a generous
1-$\sigma$ error of $\pm 0.09$, hence ${\rm (m-M)_V} = 14.47\pm 0.09$.
Adding in quadrature the error due to reddening, a total distance
modulus error of $\pm 0.10$ is adopted.  M13 has a very blue HB, so
the determination of V(HB) is difficult.  We adopt ${\rm V_{HB}} =
14.83\pm 0.10$ (\cite{cds}), implying $\mvrr = 0.36\pm 0.14$.

\subsection{Calibration of \mvrr\ via the LMC}
Walker (1992) determined the mean magnitudes of a number of RR Lyr
stars in several LMC clusters.  Adopting a distance modulus to the LMC
of $\mu_{LMC} = 18.50\pm 0.10$, he found $\mvrr = 0.44\pm 0.10$.  This
distance modulus was based upon the traditional calibration of
Cepheids.  Using Hipparcos based parallaxes, Feast \& Catchpole (1997)
derived $\mu_{LMC} = 18.70\pm 0.10$.  This distance relied
upon a period-color relation, and parallaxes of rather low quality
($\sigma_\pi/\pi \ga 0.3$).  An analysis of the Hipparcos Cepheid data by 
\cite{barry} yields $\mu_{LMC} = 18.57\pm 0.11$ who noted that ``other
effects on the Cepheid PL relation (e.g.\ reddenning, metallicity,
statistical errors) are as significant as this reassessment of its
zero point''.  The distance to the LMC may be estimated independent of
the Cepheid or RR Lyr\footnote{Reid (1997) and
\cite{gratton2} have made an estimate of the LMC distance using his
subdwarf fitting based on Hipparcos parallaxes, of $\mu_{LMC} \approx
18.65$ and $\mu_{LMC} \approx 18.60$.  However, this is based in part
on subdwarf fitting to M92 ($\feh = -2.2$).  All of the $ \feh < -2.0$
subdwarfs are either (a) evolved off the ZAMS, or (b) suspected/known
binaries, or (c) have poor \feh\ determinations.}  from geometric
considerations using the `light echo' times to the ring around SN
1987A (\cite{panagia}, \cite{gould1}).  Using the same data set, but
independent analysis, the SN1987A ring distance has been re-calculated
by a few groups. \cite{sonneborn} found $\mu_{LMC} = 18.43 \pm 0.10$,
while Gould \& Uza (1997) determined $\mu_{LMC} < 18.44 \pm
0.05$. Recently, \cite{lundq} reported a lower limit of $\mu_{LMC} <
18.67
\pm 0.08$.  In light of these contradictory results, we have elected
to follow the conclusion of \cite{barry} and adopt a distance modulus of 
$18.50$ mag for the LMC, and assume an uncertainty $\pm 0.14$ to fully
encompass the range of recently published values.
Adopting this distance modulus, along with the photometry of Walker
(1992) yields $\mvrr = 0.44\pm 0.14$ at $\feh = -1.9$.

\subsection{Theoretical HB models}
Continued advances in our understanding in the basic physics which
governs stellar evolution have lead to ever more reliable theoretical
HB models.  Recently, \cite{hbmod} have constructed synthetic HB models
for various clusters, based upon new evolutionary models for HB stars.
Assuming a primordial helium abundance of 0.23, these models predict
$\mvrr = 0.34$ for M92 ($\feh = -2.25$) and $\mvrr = 0.42$ for M15
($\feh = -2.15$).  These two clusters are among the 17 old clusters
whose mean age is determined in \S \ref{sect4}.  Taking a simple
average of the above two numbers yields $\mvrr = 0.38$ at $\feh =
-2.20$.  Also, since the primordial helium value utilized in this
analysis was on the low side, we have adjusted this mean value to 0.36
to account for a mean primordial helium value of 0.235.  We also adopt
an error of 0.10 mag on \mvrr, to allow for possible errors in the
models and in the primordial helium abundance estimate.

\subsection{Combining the Distance Estimates}
The individual determinations of \mvrr\ at the various metallicities are
summarized in Table \ref{tab3}. While it is not evident from this  
table, there
is considerable evidence from other observations and theoretical  
modeling that
\mvrr\ is a function of \feh:
\beq
\mvrr = \mu (\feh + 1.9) + \gamma.
\label{eqmvrr}
\eeq
We have chosen distance calibrations which yield reliable absolute
numbers with the minimum possible systematic uncertainties\footnote{In  
contrast to \cite{first}, we have elected not to include the
statistical parallax   results of
\cite{layden}.  This technique yields a reliable estimate of $\gamma$  
for field RR Lyr stars.  However, we now recognize that this number
may be biased compared to globular cluster RR Lyr stars.  Field RR Lyr
stars are preferentially found in places where the evolutionary
time-scales are long (ie: near the zero-age horizontal branch).  No
such selection effect exists for cluster RR Lyr stars, so one would
suspect that the typical cluster RR Lyr stars would be more evolved,
and hence, brighter than the typical field RR Lyr star.  Indeed, the
statistical parallax results lead to $\gamma = 0.62\pm 0.12$, which is
0.2 mag fainter than what we determine for cluster RR Lyr stars.}
Hence, they are useful in deriving the value of  $\gamma$.  However,  
these \mvrr\ determinations do not provide reliable information on the
\mvrr-\feh\ slope $\mu$.  For this, one needs to utilize techniques
which yield reliable relative \mvrr\ values as a function of \feh.
Theoretical HB models, as well as Baade-Wesselink studies of field RR
Lyr stars provide the best estimate of the \mvrr-\feh\
relationship\footnote{ The synthetic HB models confirm the earlier
result of \cite{Lee} that Mv(RR) depends both on [Fe/H] and HB
morphology, a plausible explanation for the different slopes derived
by different authors.  For example, if one chooses the same group of
globular clusters as \cite{gratton2}, who derived a slope of $ 0.22 \pm
0.09$ by fitting Hipparcos subdwarfs to globular cluster main
sequences, the theoretical models yield a slope near 0.25.  The
theoretical ZAMS slope is 0.20.  Note that the models show that the
evolutionary correction from the ZAMS used by \cite{carney92} should
be used with caution, as it clearly does not apply to the more extreme
HB morphologies observed in the oldest clusters.}  The semi-empirical
Baade-Wesselink method has been applied by \cite{jones}
and \cite{skillen}.  Reanalysis of these data suggest
that $\mu = 0.22\pm 0.05$ (\cite{atareview}).  The latest theoretical
models of blue HB clusters yield slopes of $\mu = 0.25\pm 0.07$
(\cite{hbmod}).  A weighted mean value of $\mu = 0.23\pm 0.04$ was adopted.
The third column in Table \ref{tab3} lists the various values of
$\gamma$ implied by the individual determinations of
\mvrr.  There is a considerable spread in these values (0.21 to 0.52),
reinforcing the notion that the dominant uncertainty remains
systematic and that our previous procedure of assigning a Gaussian
uncertainty to this quantity was ill-advised.  We have thus now chosen
to utilize a uniform top-hat uniform distribution which evenly weights
all values in the range 0.21 to 0.52.  However to give some emphasis
to the mean value of the measured data, we have added to this
distribution a Gaussian distribution centered on the weighted mean of
$\gamma = 0.39$ with an uncertainty of 0.08,
doubling the calculated error in the mean to account for the average
deviation from the mean.  Note that this new mean value is 0.21 mag
(more than $2-\sigma$) below the value adopted in \cite{first}.  This
will lead to a considerably downward revision in our GC age estimates,
which we believe will also now have a distribution which is more
appropriate to the systematic nature of the existing uncertainties.
(We emphasize once again that while the Hipparcos data provided 
a motivation for re-examining this value, 
all of the other distance estimators 
we have examined 
apear now to be consistent, within the systematic 
uncertainties quoted, with a much 
lower value than we previously adopted.)

Spectroscopic studies of blue horizontal-branch (BHB) stars provide
further support for the longer GC distance scale adopted here. 
From both the continuous spectrum and
absorption line profiles, it is possible, with the help of model stellar
atmospheres, to derive the effective temperature and surface gravities
of these stars.  This combined information yields the mass-to-light
ratio M/L of the star, and if its distance is also known, its
mass. This method has in the past yielded masses incompatible with the
standard HB evolution theory (masses lower than evolutionary
models)(\cite{deBoer95}; \cite{moehler},1997).  A recent
attempt to rederive the distances of some field BHB stars using 
Hipparcos parallaxes, could not be given much weight in view of the
smallness of the parallaxes (\cite{deBoer97}).  Using Reid's
(1997) reanalysis of the distances to some globular clusters based on
larger Hipparcos subdwarf parallaxes, \cite{heber} have
reconsidered this problem, and concluded that the higher luminosities
for BHB stars now yield masses in better agreement with the
evolutionary masses.  This important result provides independent
support, based on physical modeling, for revising upward the
distance scale to globular clusters, as suggested by several lines of
reasoning, including the Hipparcos parallax data.

Finally, we should point out that the new distance scale yields a poor
fit to calculated isochrones near the cluster turnoff.  This suggests
that the stellar model radii may need revision (better stellar
atmospheres and convection modeling).  Improvements in atmosphere
models may lead to revisions of the $T_{eff}$ to color
transformations, particularly for the most metal poor stars.  Since
the \dv\ method is little affected by surface effects, it further
justifies our preference for this approach over the \dbv\ approach,
and the approach of fitting to the shapes of theoretical isochrone
turnoffs, both of which are sensitively affected by atmosphere and
outer envelope physics.

\section{The Monte Carlo Variables \label{sect3}}
In order to access the range of error associated with stellar
evolution calculations and age determinations, the various inputs into
the stellar evolution codes were varied within their uncertainties.
In this Monte Carlo analysis, the input parameters were selected
randomly from a given distribution.  The distributions are based on a
careful analysis of the recent literature, as summarized in
\cite{first}.  As ages will be derived using \mvto\ most attention was
paid to parameters which could effect the age-\mvto\ relationship.
Table \ref{tab4} provides an outline of the various input parameters
and their distribution.  If the distribution is given as statistical
(stat.), then the parameter in question was drawn from a Gaussian with
the stated $\sigma$.  If the error was determined to be a possible
systematic (syst.) one, then the parameter was drawn from a top-hat
(uniform) distribution.  In total, 1000 independent sets of isochrones
were calculated.  Each set of isochrones consisted of three different
metallicities ($\feh = -2.5, -2.0$ and $-1.5$) at 15 different ages 
($8 - 22$ Gyr)  (see \cite{first} for further details).

\section{Results \label{sect4}}
\subsection{The Technique}
The absolute magnitude of the main sequence turn-off is the favored
age determination technique when absolute stellar ages are of interest
(see discussion in \cite{first}). Turn-off luminosity ages can be
determined independent of reddening by using the difference in
magnitude between the main sequence turn-off and the HB, \dv.  Each
set of Monte Carlo isochrones provides an independent calibration of
\mvto\ as a function of age.  This was combined with the \mvrr\
calibration discussed in \S \ref{sect2} to determine a grid of
predicted \dv\ values as a function of age and \feh\ which
is then fit to an equation of the form
\beq
t_9 = \beta_0 + \beta_1\dvtwo + \beta_2\dvtwo^2 + \beta_3\feh +
\beta_4\feh^2 + \beta_5\dvtwo\feh,
\label{fit}
\eeq
where $t_9$ is the age in Gyr.  The observed values of \dv\ and \feh,
along with their corresponding errors, are input in (\ref{fit}) to
determine the age and its error for each GC in our sample.

The age determination for any individual globular cluster has a large
uncertainty, due to the large observational errors in ${\rm V}$(TO).
This error is minimized by determining the mean age of a number of
globular clusters.  However, there is a significant age range among
the globular clusters (e.g.\ \cite{sard}, \cite{van2},\cite{buonanno},
\cite{cds}).  This problem was avoided by selecting a sample of 
globular clusters which are well observed, metal-poor ($\feh \le
-1.6$), and which are not known to be young (based on HB morphology
and/or the difference in color between the turn-off and giant
branch).  In the tabulation of \cite{cds}, 17 GCs satisfy the above
criteria: NGC 1904, 2298, 5024, 5053, 5466, 5897, 6101, 6205, 6254,
6341, 6397, 6535, 6809, 7078, 7099, 7492, and Terzan 8.  The
observational data for each cluster was taken from \cite{cds}.  The
mean (and median) metallicity of this sample is $\feh = -1.9$.

\subsection{A Likelihood Distribution for the Age of the Oldest  
Globular Clusters\label{sectage}} 
To derive our best estimate for the age, and uncertainty in the age of
the oldest GCs, a mean age and $1\sigma$ uncertainty in the mean was
determined for each set of isochrones, and a given value of
\mvrr.  The value of \mvrr\ was taken to be a random variable, weighted  
as described earlier (a top hat distribution between 0.21 and 0.52
superimposed on a Gaussian distribution with mean and uncertainty
$\mvrr = 0.39\pm 0.08$), and the sets of isochrones were sampled with
replacement 12,000 times.  For each sample, we recorded a random age
drawn from a Gaussian distribution with the mean age and variance for
that isochrone set at that value of \mvrr .    

\begin{figure}
\centerline{\psfig{figure=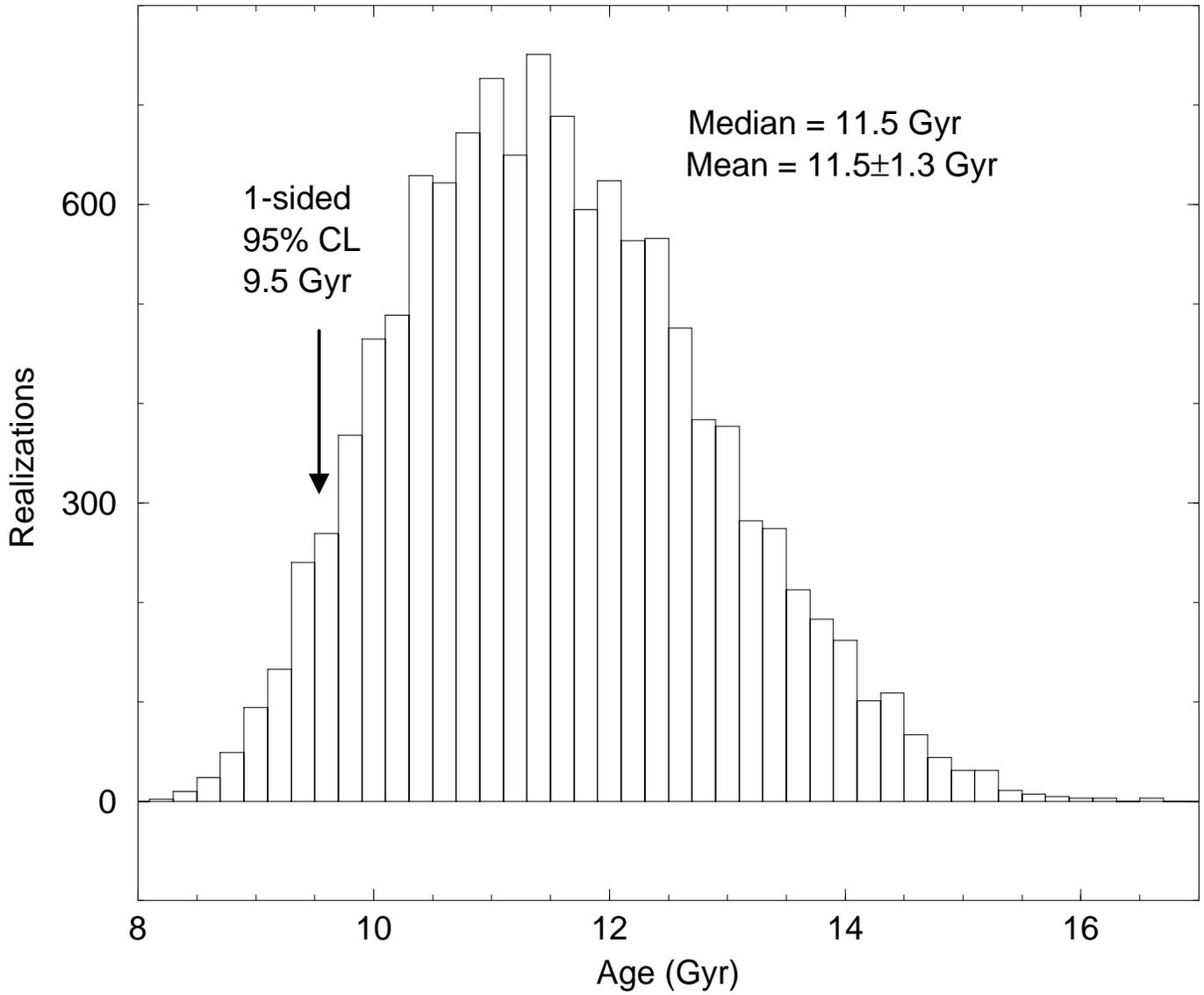,height=18.0cm,angle=270}  }
\caption{Histogram of GC ages.  The median, mean, standard deviation
and one-sided, 95\% confidence level lower limit are all indicated on
the figure.}
\label{fighist}
\end{figure}

The age data were sorted and binned, to produce the histogram shown in
Figure \ref{fighist}.  The median and mean age is 11.5 Gyr, with a
standard deviation of 1.3 Gyr.  The 1-sided, 95\% lower confidence
limit is 9.5 Gyr, and is believed to represent a robust lower limit to
the age of the GCs, and more properly takes into account the residual
systematic uncertainties in \mvrr, which largely determine the width
of the derived age distribution.  We are fully aware that due to our
revision of the \mvrr\ zero-point, these ages are considerably reduced
compared to the ages given in \cite{first}.  Indeed, our new mean age
is below our previous claimed $95\% $ lower limit which was based on
the assumption of Gaussian uncertainty in \mvrr.  In any case, our new
results considerably alter the constraints one can derive on
cosmological models (see \S \ref{sect5}).

Even though we have considered four independent distance
determinations in addition to the Hipparcos parallaxes, our age
estimate is in good agreement with two recent works which relied
solely on Hipparcos parallaxes to determine the distances (and hence,
ages) to a number of GCs (\cite{reid}; \cite{gratton2}).  \cite{pont}
have determined an age of 14 Gyr for M92, which is in disagreement
with our work.  \cite{pont} made a new fit of the CMD of M92 to theoretical
isochrones, based on the Hipparcos subdwarf data.  This paper represents
a comprehensive analysis of the available data, and attempts
the difficult task of correcting for selection effects which are more
relevant in this case than the classical \cite{lutzk} corrections.  
However, our own analysis suggests that they have overestimated the
corrections to the Hipparcos parallaxes due to biases (see Appendix).
Their corrections due to the presence of binaries is very uncertain
($\pm 0.15\,$mag); a fact which was not considered by \cite{pont} in
their analysis.  A better procedure, which is not to include the
suspected binaries in the fit, yields a larger distance modulus for
M92.  This approach, as pointed out by \cite{pont} yields ${\rm
(m-M)_V} = 14.74\pm 0.08$ mag.  With this distance modulus, and the
photometry of \cite{m92} (the same photometry used by \cite{pont}), we
calculate that the absolute magnitude of the point on the subgiant
branch which is 0.05 mag redder than the turn-off is ${\rm V(BTO)} = 3.39\pm
0.08$ mag.  This point is an excellant diagnostic of the absolute age
of M92 (\cite{paper2}), and using our isochrones (as outlined in 
\S \ref{sectmvbto}) results in an age for M92 of $12.1\pm 1.3\,$Gyr.
This is in good agreement with our estimate for the mean age of the
oldest GCs (which includes M92) given above.

In the final analysis, the \cite{pont} paper puts most of the weight
of their fit on the agreement between the shapes of the theoretical
isochrones and the data near the turnoff.  However, this optimistic
assessment of the models does not seem warranted in view of the
well-known uncertainties associated with the treatment of convection,
and the neglect of diffusion in the isochrones used (helioseismoly has
taught us that diffusion must be taken into account in the Sun
(\cite{basu}, \cite{guenther}).  The need to apply an arbitrary color
shift to the \cite{vand3} isochrones to reproduce the observed colors
of M92, is another indication of the uncertainties involved, and lends
further support to the choice of the \dv\ method in dating globular
clusters.  We conclude that taking into account the differences in
adopted distance moduli, and the neglect of diffusion by \cite{vand3},
our age estimate for M92, which is $11.5\pm 1.3$ Gyr, is in good
agreement with the Hipparcos data presented by \cite{pont}.

\subsection{Effect of \mvrr\ on the age estimate}
\begin{figure}
\centerline{\psfig{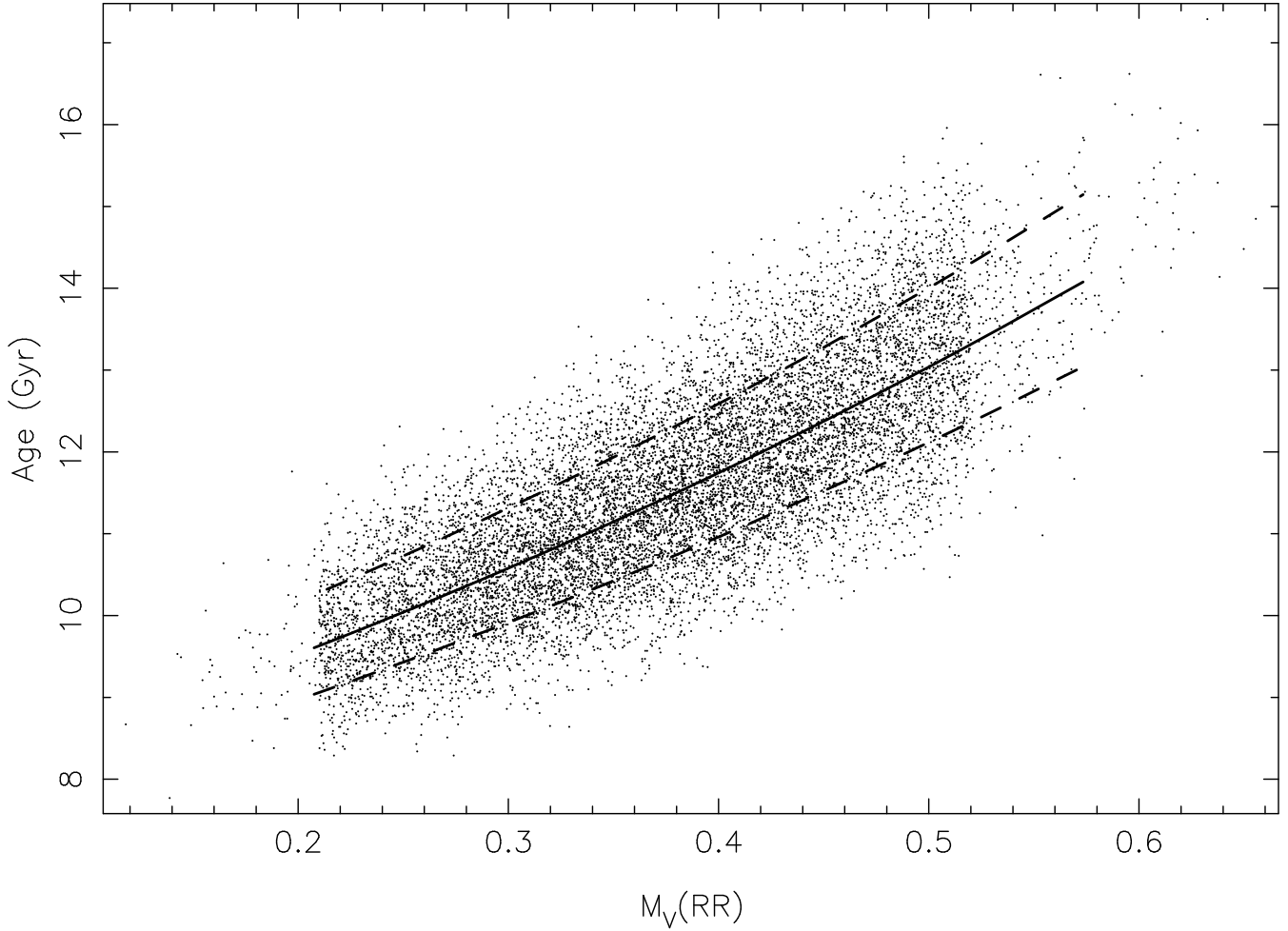}  }
\caption{Best estimate for the age of the oldest GCs, as a function of
the adopted \mvrr\ value at $\feh = -1.9$. The best fitting median
(solid line), along with $\pm 1\,\sigma$ limits (dashed lines) are
plotted.  These lines are log fits ($\log (t_9) = a + b\,\mvrr$), with
the following cofficients: median 
$(a,b) = (  0.888,\;   0.454)$; 
$-1\,\sigma\, (a,b)= (  0.866,\;   0.436)$; and 
$+1\,\sigma\, (a,b)= (  0.915,\;   0.463)$.
Further details are provided in the text.  }
\label{figmvrr}
\end{figure}

As was emphasized in \cite{first} (and by other authors), the principal
uncertainty in absolute GC age determinations is the distance scale.
With the \dv\ age determination technique, this translates into the
uncertainty in \mvrr.  We explicitly display this effect in Figure
\ref{figmvrr}, where the GC ages are plotted as a function of \mvrr.  
In order to quantify this uncertainty, median and $\pm 1\,\sigma$
points were determined as a function of \mvrr.  These were obtained by
sorting the data based on \mvrr, and then binning the ages as a
function of \mvrr.  Sixty bins (corresponding to 200 ages per bin)
were used, and in each bin the median age, and $\pm 1\,\sigma$ (68\%
range) ages were determined.  An inspection of these points revealed
that a simple linear relationship existed when one used the log of the
age. A linear function of the form $\log (t_9) = a + b\,\mvrr$ was
fitted to this data, and the coefficients of this fit are given in the
figure caption.

The median and $\pm 1\,\sigma$ fits are extremely useful summaries of
our result.  
For example, at $\mvrr = 0.40$, the median fit yields  11.7 Gyr,
identical to that   given by the entire distribution (Fig.\
\ref{fighist}).  The $\pm 1\,\sigma$ fits yield ages of 12.6 and 11.0
Gyr.  Thus, if \mvrr\ was known to be exactly 0.40, then the error in
the age of the oldest GCs would be $\pm 0.8$ Gyr, due solely to the
residual theoretical uncertainties in the stellar evolution
calculations.  The median and $\pm 1\,\sigma$ fits we present here may
be used to update our age estimate as further data are obtained.  For
example, if $\mvrr = 0.50\pm 0.05$ then from the fits, the median age
would be $13.03\,$Gyr, with an error of $\pm 0.94\,$Gyr due to the
theoretical uncertainties aside from those associated with \mvrr.
Next, from the median fit, the median age at $\mvrr =0.45$ and $0.55$
may be determined (corresponding to $\pm 0.05\,$mag) in order to
estimate that the uncertainty in age associated with the
\mvrr\ uncertainty is $\pm 0.68$ Gyr.  Combining these two error  
estimates in quadrature ($\pm 0.94$ and $\pm 0.68$) would result in a best  
estimate of $13.1\pm 1.2$ Gyr for $\mvrr = 0.50\pm 0.05$.  To verify this
result, we have re-run the Monte Carlo analysis with the above choice
of \mvrr\ and found identical results to those obtained from the
\mvrr\ median and $\pm 1\,\sigma$ fits above.

\subsection{Effect of the stellar evolution parameters on the age  
estimate}
In order to examine how the individual stellar evolution parameters
(given in Table \ref{tab4}) affect the estimated age, the mean age of
the 17 GCs was determined for each of the 1000 Monte Carlo isochrones
assuming fixed value $\mvrr = 0.40$. In a procedure analogous to that
used for the \mvrr\ fits, median and $\pm 1\,\sigma$ fits were
determined for each of the 13 continuous variables listed in Table
\ref{tab4}.  As only 1000 points were available, only 20 bins were
used.  In addition, it was found that (due to the reduced age range),
a linear fit provided as good a description as a log fit.  Thus, the
median and $\pm 1\,\sigma$ fits for each parameter $x$, were of the
form $t_9 = a + b\,x$.

\begin{figure}
\centerline{\psfig{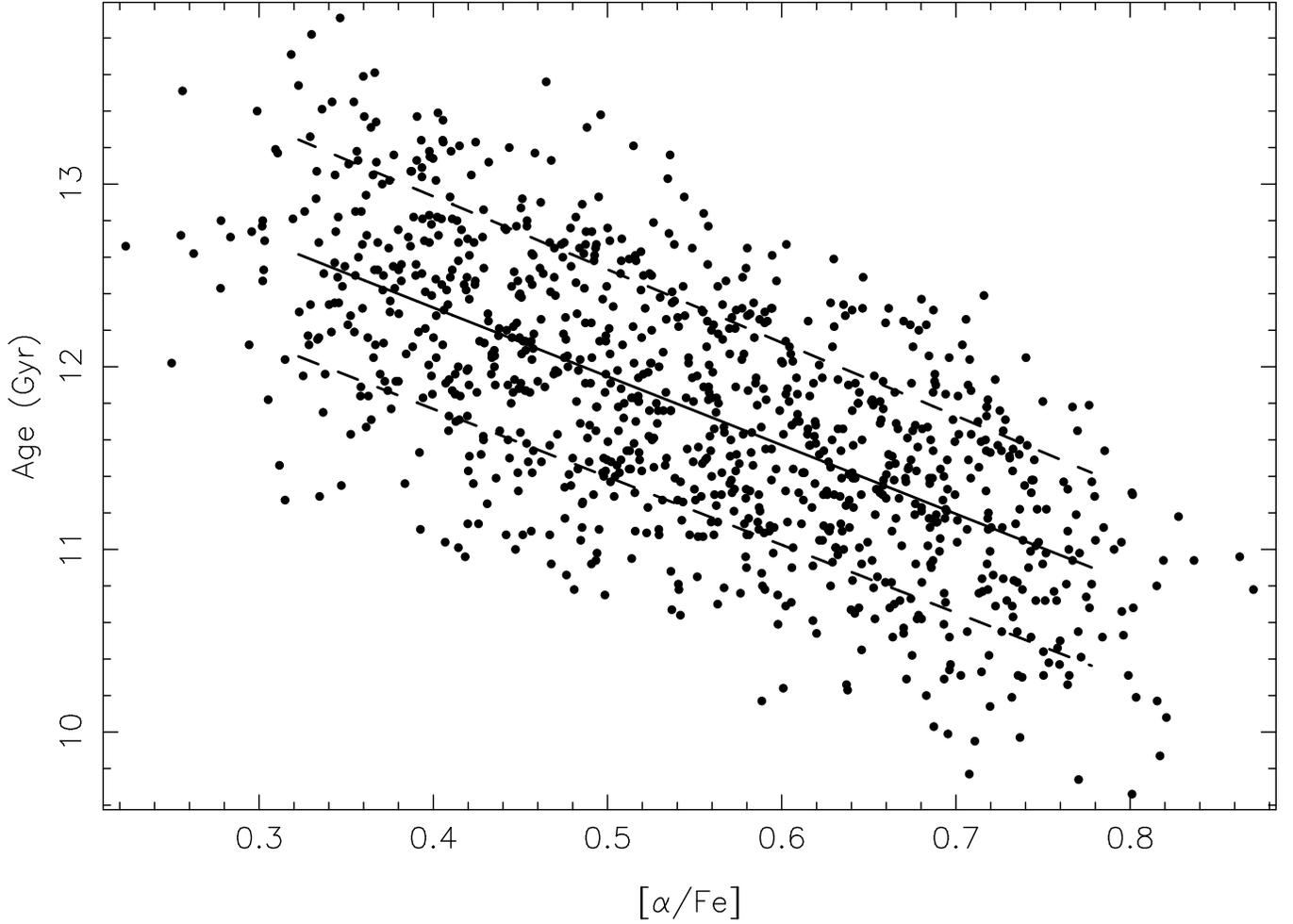}  }
\caption{Age as a function of $[\alpha/{\rm Fe}]$.
The best fitting median,
along with $\pm 1\,\sigma$ limits are plotted.  These lines are of the  
form
$t_9 = a + b\,[\alpha/{\rm Fe}]$, with the following cofficients:
median  $(a,b) = ( 13.83,\;  -3.77)$; 
  $-1\,\sigma\, (a,b)= ( 13.26,\;  -3.72)$; and 
  $+1\,\sigma\, (a,b)= ( 14.54,\;  -4.00)$.
}
\label{figalpha}
\end{figure}

\begin{figure}
\centerline{\psfig{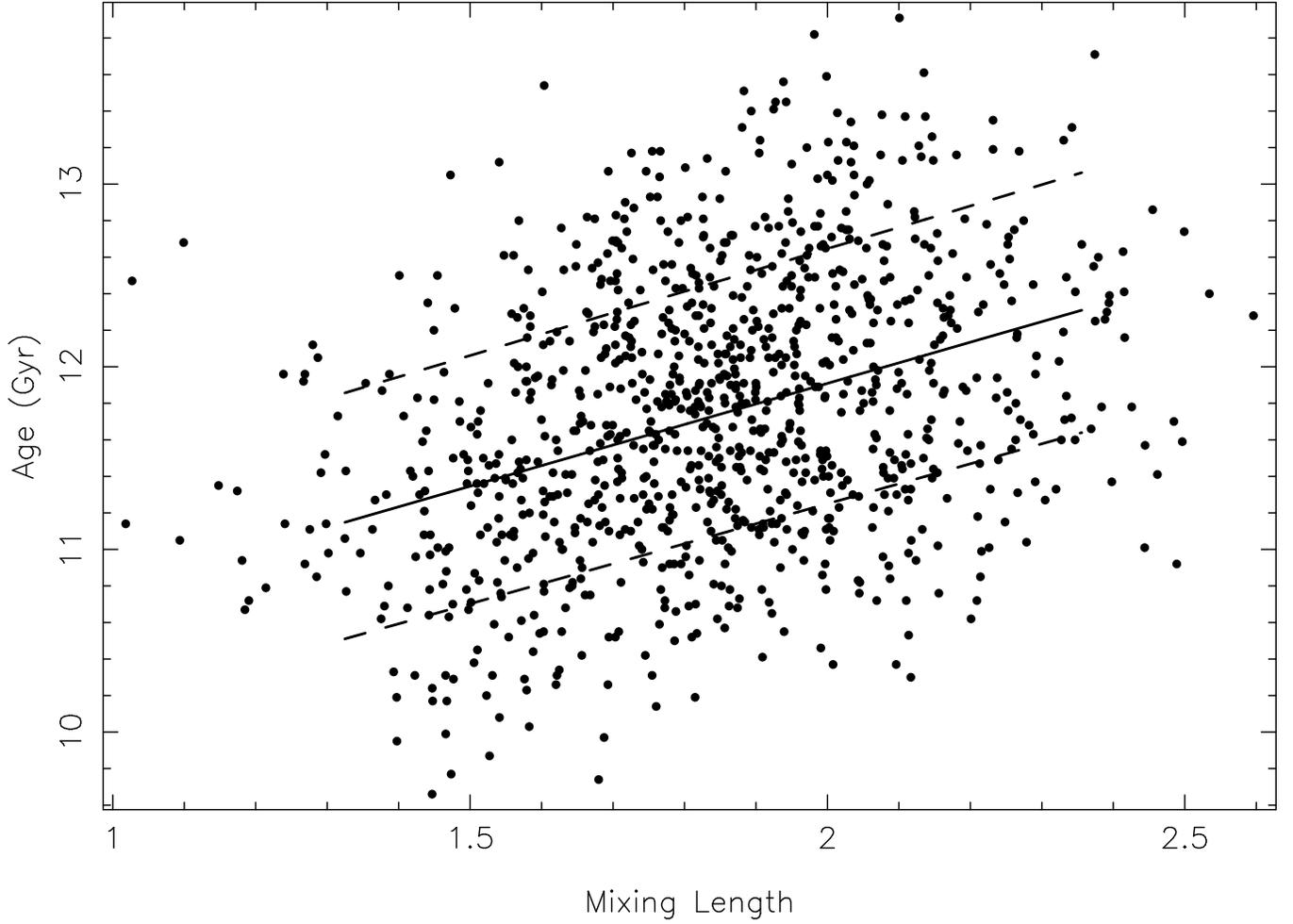}  }
\caption{Age as a function of the mixing length ($\alpha$) used in 
the stellar  
models.  The lines of the form $t_9 = a + b\,\alpha$ have the
following cofficients: median $(a,b) = ( 9.66,\; 1.13)$; $-1\,\sigma\,
(a,b)= ( 9.06,\; 1.10)$; and $+1\,\sigma\, (a,b)= ( 10.31,\; 1.17)$.
}
\label{figmix}
\end{figure}

\begin{figure}
\centerline{\psfig{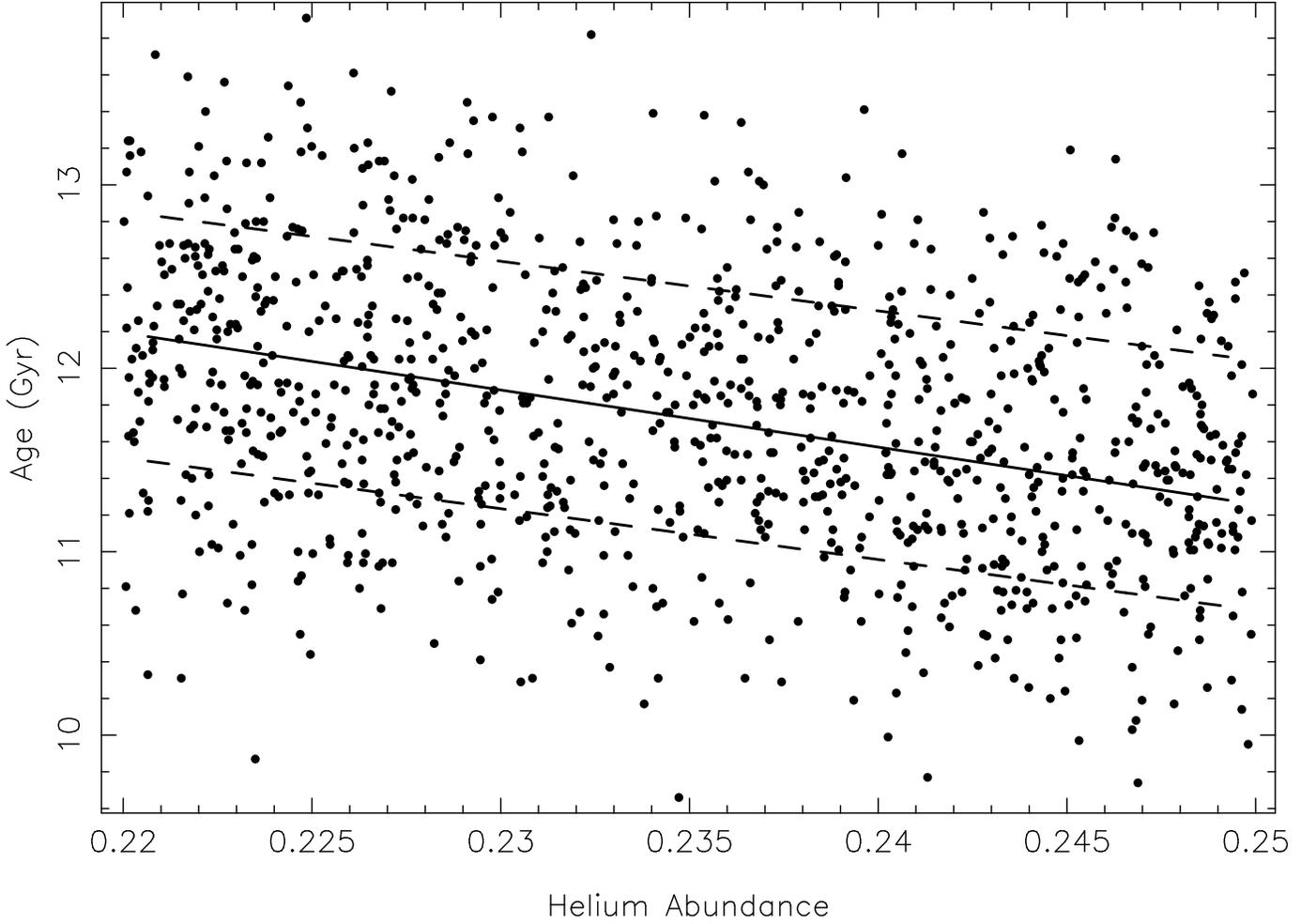}  }
\caption{Age as a function of the helium abundance ($Y$) used 
in the stellar models.    
The lines of the form $t_9 = a + b\,Y$ have the 
following cofficients: median $(a,b) = ( 19.05,\; -31.15)$; 
  $-1\,\sigma\, (a,b)= ( 17.61,\; -27.73)$; and 
  $+1\,\sigma\, (a,b)= ( 18.80,\; -27.04)$. 
}
\label{fighelium}
\end{figure}

\begin{figure}
\centerline{\psfig{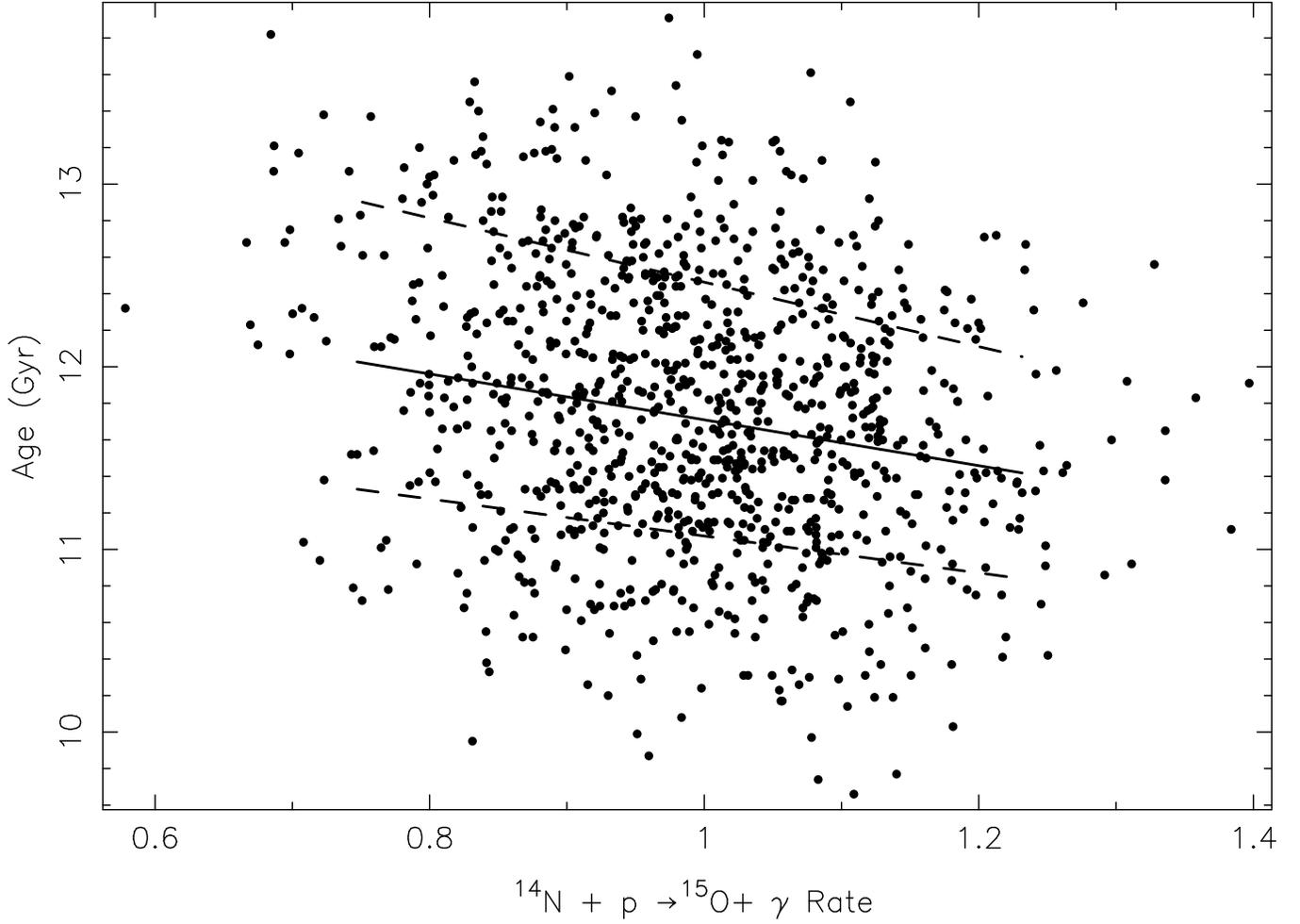}  }
\caption{Age as a function of the ${\rm {^{14}N} + p \longrightarrow
{^{15}O}} + \gamma$ reaction rate ($\Re$).  
The lines of the form $t_9 = a + b\,\Re$ have the 
following cofficients: median $(a,b) = ( 12.97,\;  -1.26)$; 
  $-1\,\sigma\, (a,b)= ( 12.09,\;  -1.01)$; and 
  $+1\,\sigma\, (a,b)= ( 14.22,\;  -1.76)$.
}
\label{fignuc}
\end{figure}

\begin{figure}
\centerline{\psfig{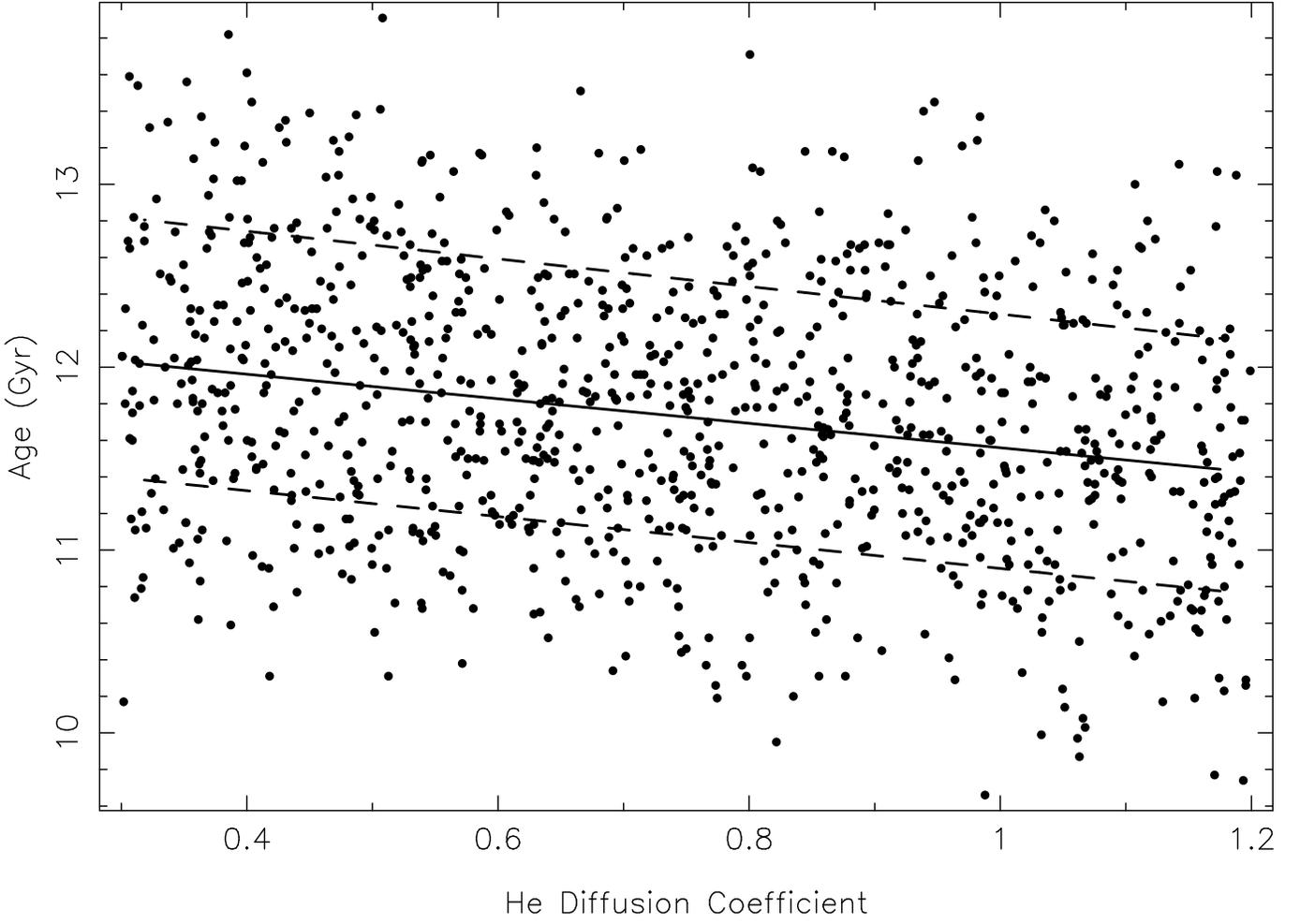}  }
\caption{Age as a function of the helium diffusion coefficient ($D$) used 
in the stellar models.    
The lines of the form $t_9 = a + b\,D$ have the 
following cofficients: median  $(a,b) = ( 12.23,\;  -0.67)$; 
  $-1\,\sigma\, (a,b)= ( 11.61,\;  -0.71)$; and 
  $+1\,\sigma\, (a,b)= ( 13.05,\;  -0.76)$.
}
\label{figdiff}
\end{figure}

\begin{figure}
\centerline{\psfig{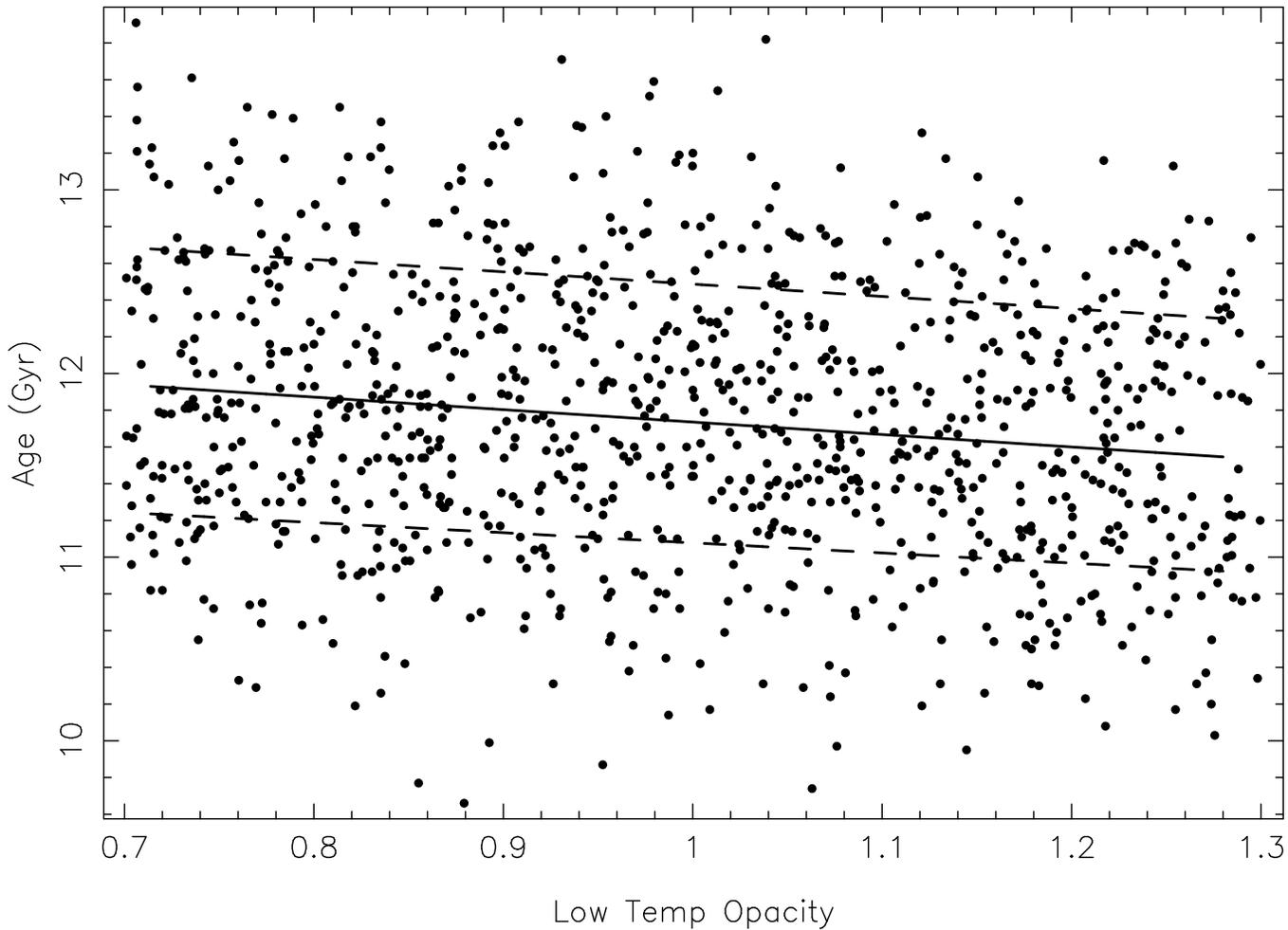}  }
\caption{Age as a function of low temperature opacity ($\kappa$).  
The best fitting median,
along with $\pm 1\,\sigma$ limits are plotted.  These lines are of the  
form $t_9 = a + b\,\kappa$, with the following cofficients:
median $(a,b) = ( 12.41,\;  -0.68)$; 
  $-1\,\sigma\, (a,b)= ( 11.63,\;  -0.55)$; and 
  $+1\,\sigma\, (a,b)= ( 13.16,\;  -0.67)$.
}
\label{figlow}
\end{figure}

This procedure revealed that several of the input parameters had a
negligible effect on the derived ages of the globular clusters.  In
order of importance, the following parameters were found to impact the
GC age estimate: $\alpha/[{\rm Fe}]$, mixing length, helium abundance,
${\rm {^{14}N} + p
\longrightarrow
{^{15}O}} + \gamma$ reaction rate, helium diffusion coefficient, and  
the low temperature opacities.  The plots of age as a function of these
important parameters are shown in Figures \ref{figalpha} ---
\ref{figlow}.   The figure captions give the coefficients of the median
and $\pm 1\,\sigma$ fits for each of the variables.  These fits can be
used to update our best estimate for the age of the oldest globular
clusters (in a manner analogous to that described for the \mvrr\
fits), as improved determinations of the above quantities become
available.

\begin{figure}
\centerline{\psfig{figure=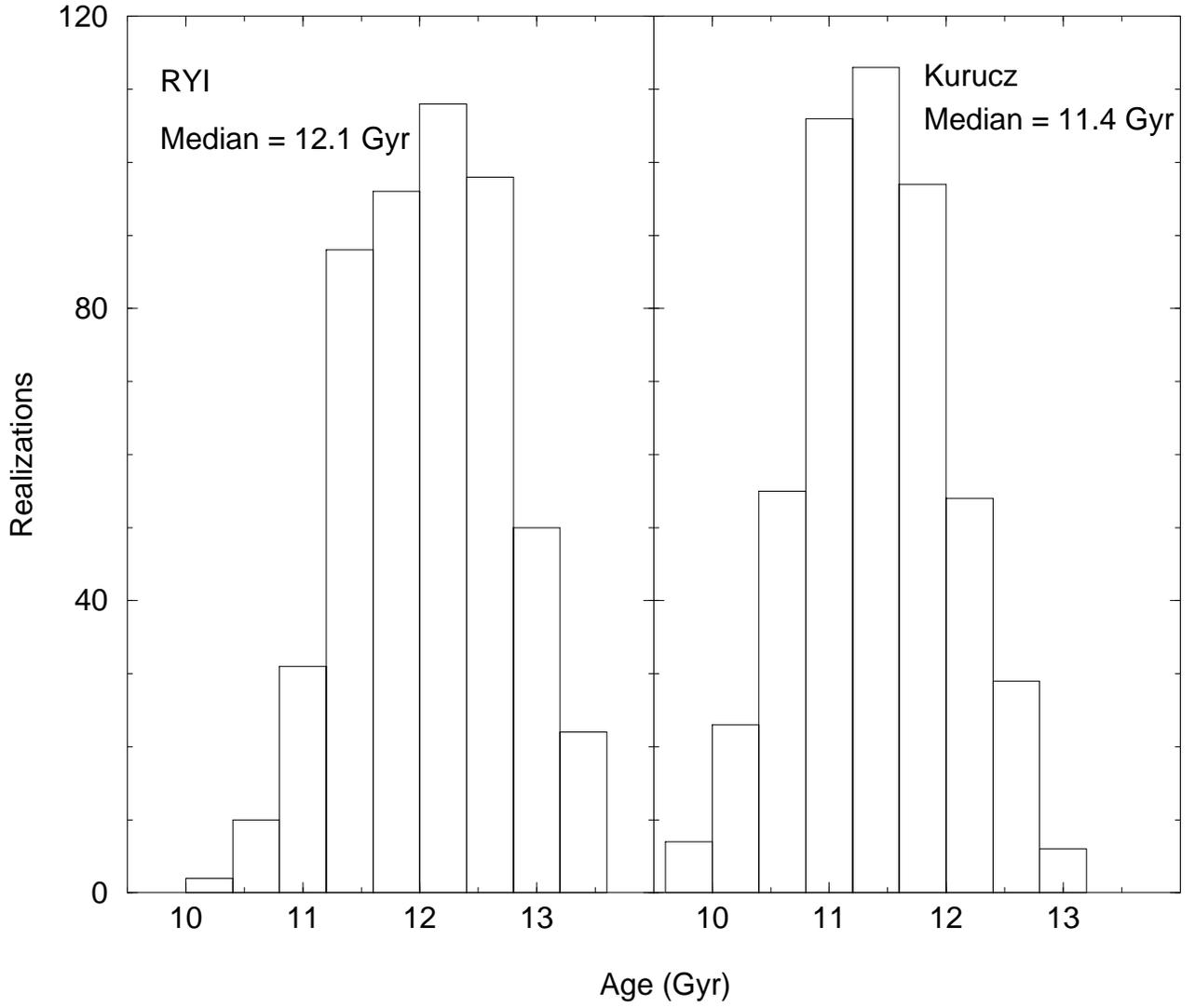,height=15.5cm,angle=270}  }
\caption{Histograms for the mean age of the oldest globular clusters,
using (a) the  RYI (Green \ea\ 1987) colour table, or (b) the Kurucz  
1992 colour table. 
}
\label{figcoltab}
\end{figure}

In addition to the 13 continuous variables, we considered two binary
variables (surface boundary condition, and color table, see Table
\ref{tab4}).  To examine the effect these parameters have on the
derived ages, the ages were divided into 2 groups depending on which 
 surface boundary condition (color table) was used in the stellar
evolution codes.  Histograms  were constructed for each group, and
compared.  Not surprisingly, we found that the choice of the surface
boundary condition had a negligible impact on the derived ages.
However, the choice of the color table was important, and the two
histograms are plotted in Figure \ref{figcoltab}.  The choice of the
color table changes the median age by 0.7  Gyr.

\subsection{Calibration of the \mvto\ and \mvbto\ age relations 
\label{sectmvbto}}
If the distance modulus to some cluster is known, then an accurate
absolute age may be determined using \mvto\, or alternatively using
\mvbto\
(\cite{paper2}). This later point is defined to be the point on the  
subgiant
branch which is 0.05 mag redder (in B---V) than the turn-off.  As we  
have discussed, this point is easy to measure on an observed color
magnitude diagram, yet has similar theoretical uncertainties  to
\mvto\ (\cite{paper2}). As a result, the precision in age estimation  
for individual clusters is better using \mvbto.  The Monte
Carlo isochrones may be used to quantify the error associated with an
age determined via either method.  To facilitate such error estimates,
we have calculated the median and $\pm 1\,\sigma$ \mvbto\ (\mvto)
points as a function of age (in a manner similar to that described in
the previous subsection) for four values of \feh: $-2.5\; -2.0\; -1.5$
and $-1.0$.  For ages between 8 and 17 Gyr, these points were then fit
to a function of the form
\beq
\log(t_9) = \beta_1 + \beta_2{\rm M_V} + \beta_3\feh  
+\beta_4\feh^2
+ \beta_5\feh\,{\rm M_V}
\label{eqmvbto}
\eeq
where ${\rm M_V}$ was chosen to be either \mvbto\ or \mvto.
The coefficients of the median and $\pm 1\,\sigma$ fits, for both
\mvbto\ and \mvto\ are given in Table \ref{tab5}.

The use of these fits for determining ages via \mvbto\ is illustrated  
for NGC 6752.  Averaging the white dwarf distance modulus (\S
\ref{white}) and  the subdwarf main sequence fitting modulus (\S
\ref{sectn6752})  results in ${\rm (m - M)_O} =13.12\pm 0.07$, or 
${\rm (m - M)_V} =13.25\pm 0.07$.  Using the photometry of
\cite{penny}, we find ${\rm V(BTO)} = 16.83\pm 0.04$, so that ${\rm
M_V(BTO)} = 3.58\pm 0.08$. Recall that $\feh = -1.51\pm 0.08$ (\S
\ref{sectn6752}).  Using the coefficients of the fits in Table 1, this
corresponds to an age of $11.15\pm 0.8\,$Gyr if \mvbto\ and \feh\ were
known exactly.  The effects of the \mvbto\ and \feh\ errors may be
taken into account by using the median fit, and calculating ages for
the $\pm 1\,\sigma$ values for \mvbto\ and \feh.  This procedure
results in estimated errors of $\pm 0.9\,$Gyr due to the \mvbto\ error
($\pm 0.08\,$mag) and $\pm 0.4\,$Gyr due to the \feh\ error of $\pm
0.08\,$dex.  Adding all three errors together in quadrature yields an
age of $11.2\pm 1.3\,$Gyr for NGC 6752.  This intermediate metallicity
cluster has an age quite similar to the mean age of the 17 metal-poor
clusters ($11.5\pm 1.3\,$Gyr) determined in \S \ref{sectage}.  More
important, note that the uncertainty on the age of NGC 6752 determined
in this way is comparable to the uncertainty in the mean of the set of
17 old globular clusters, illustrating the potential power of the
method based on \mvbto.

Similarly, for M5, we calculate ${\rm V(BTO)} = 18.03\pm 0.02$ using  
the photometry of \cite{sand}.  With ${\rm (m-M)_V} = 14.51 \pm0.09$ 
(\S \ref{sectm5}), this results in $\mvbto = 3.52\pm 0.09$.  Assuming
$\feh = -1.17\pm 0.08$ (\cite{snedenm5}), and using the technique
outlined for NGC 6752, an age of $8.9\pm 1.1\,$Gyr is derived.
Finally, for M13 with $\feh = -1.58$ and ${\rm (m-M)_V} = 14.47\pm 0.09$ (\S
\ref{sectm13}) we find  ${\rm V(BTO)} = 18.00\pm 0.04$ using the
photometry\footnote{The deep photometry of \cite{richer} used in the
main sequence fitting (\S \ref{sectm13}) contains very few subgiant
stars, and so does not lead to a precise V(BTO) value.  The photometry
of \cite{van2} used here appears to be on the same system as the 
\cite{richer} photometry.} of \cite{van2}, resulting in 
$\mvbto = 3.53\pm 0.10$ and an age of $10.9\pm 1.4\,$Gyr.  Our results
for the distances and ages of these three clusters are summarized in
Table \ref{tab6}.



\section{Summary \label{sect5}}

Our new work has two primary results.  First, we have updated the
absolute age estimate, and quantified the uncertainty in this estimate
for the oldest globular cluster mean age.  This update is primarily
due to a reanalysis of estimates for the quantity which dominates the
age uncertainty: \mvrr\ (the distance scale to GCs).  We have
concentrated on exploring in detail different estimates in order to
account for the mean value, and the distribution in the uncertainty of
this quantity.  We find that all the data, not merely the
recent Hipparcos parallax measurements, 
suggests a large systematic
shift in \mvrr\ of approximately $0.2$ magnitudes compared to 
earlier estimates.  This has the
effect of reducing the mean age of the oldest globular clusters by
almost 3 Gyr.  At the same time, this new data makes it clearer that
\mvrr\ residual uncertainties are primarily systematic, reminding us
that even apparently gaussianly distributed measurements in
astrophysics may be subject to large systematic shifts.  As a result,
we now incorporate a large systematic uncertainty in the claimed mean
value of \mvrr\ in our estimates.  

Next, we provide a formalism which may be used by other researchers to
update the estimates given here as new data emerges.  In particular,
we have presented an explicit discussion of the effect of other input
parameter uncertainties from stellar evolution theory on the inferred
GC ages estimates.  We have displayed these effects in Figures 2-9,
and provided analytical fits for both median ages, and uncertainties
in age as a function of these parameters, and also as a function of
\mvrr.

We have also explicitly provided the fit for individual globular
cluster ages and uncertainties as a function of metallicity and
turn-off magnitude, using both the \mvto\ and \mvbto\ schemes.  This
should allow one to derive the age, and uncertainty in age for any GC
with $ -2.5 \le \feh \le -1.0$.  We have illustrated this scheme, for
the \mvbto\ method for NGC 6752, using the average distance modulus
from white dwarf sequence fitting, and subdwarf main sequence fitting,
yielding an age of $ 11.2 \pm 1.2\,$Gyr, illustrating that the \mvbto\
method in principle allows an age precision on individual GC age
determinations comparable to the \mvto\ method applied to the ensemble
of 17 old Globular clusters used in our analysis.

Finally, we briefly comment here on the cosmological implications of
our central result that the mean of 17 old, metal-poor GC is $11.5\pm
1.3\,$Gyr, with a one-sided, 95\% confidence level lower bound of 9.5
Gyr (see Krauss (1997) for further details).  
First and foremost, this results suggests that the long-standing
conflict between the Hubble age, and GC age estimates for a flat
matter dominated universe is now resolved for a realistic range of
Hubble constants.  A flat universe has an age which exceeds our lower
limit on the GC ages for a Hubble constant $H_0 \le 67\,\kmsmpc$, 
which is well within the range of current measured values.
Thus, it now appears that the ``age problem" is now no longer the
primary motivation for considering a non-zero cosmological constant in
the universe (i.e. \cite{Kraussturn}), and requires an alteration in
the arguments associated with the debate between an open, flat matter
dominated, and flat cosmological constant cosmologies 
(\cite{krauss}).

If measurements of the Hubble constant continue to converge on the
range $60 -70\,\kmsmpc$, as suggested by the most recent
analyses, cosmological concordance, at least as far as age is
concerned, will perhaps for the first time be possible in all three
scenarios.

\acknowledgments 

We would like to thank Bill van Altena, Sidney van den Bergh and the
anonymous referee whose comments have significantly improved
the final product as presented here.  This
research has made use of the SIMBAD database, operated at CDS,
Strasbourg, France and data obtained from the ESA Hipparcos Astrometry
Satellite.  BC was supported for this work by NASA through Hubble
Fellowship grant number HF--01080.01--96A awarded by the Space
Telescope Science Institute, which is operated by the Association of
Universities for Research in Astronomy, Inc., for NASA under contract
NAS 5--26555. LMK was supported in part by funds from CERN, CWRU, and
a grant from the DOE.

\appendix
\section{Potential Biases in the Subdwarf Sample \label{appen1}}
The parallaxes and absolute magnitudes for the stars listed in Table
\ref{tab2} do not include any statistical correction for possible
biases in the sample.  There are a few sources of potential biases in
the sample.  The classical \cite{lutzk} correction is a statistical
correction which takes into account systematic effects due to the fact
that (a) stars with parallaxes measured too high have a higher
probability of being included in the sample than those with parallaxes
measured too low (due to our $\sigma_\pi/\pi < 0.1$ selection
criterion), and (b) more weight is given to stars with parallaxes that
are overestimated rather than to stars with underestimated parallaxes
(due to our use of a weighted least squares fit).  In addition to
this, \cite{pont} point out that the since metal-poor stars are far
less numerous than more metal-rich stars, there may be an average
underestimation of \feh\ in the sample.  The importance of these
biases will depend on the selection criterion which are used to select
the subdwarfs used in the main sequence fitting.  The 3 papers which
have used Hipparcos subdwarf parallaxes to determine GC distances have
all had different selection criterion, and have determined different
bias corrections.  In their study, \cite{pont} determined that the
unevolved subdwarfs had a mean bias of $+0.64$ mag.  In contrast,
\cite{gratton2} determined a bias correction of $-0.004$ mag.
\cite{reid} whose subdwarf sample consisted of high proper motion
stars, elected to use individual Lutz-Kelker corrections, whose
magnitude depended on the uncertainty in the parallax.  In general,
the corrections used by \cite{reid} were small, and in the opposite
sense to those employed by \cite{pont}.

Our subdwarf study differs from the those of \cite{reid},
\cite{gratton2} and \cite{pont} in that we have access to the entire
Hipparcos catalogue.  Stars were selected for inclusion in the
Hipparcos input catalogue based on a variety of considerations, and so
there is no well defined selection criterion for the entire Hipparcos
catalogue.  Thus,  it is difficult to assess the importance of the
various biases {\it a priori}.  For this reason, we have elected to
use a stringent selection criterion $\sigma_\pi/\pi < 0.1$
which minimizes the importance of the 
\cite{lutzk} type bias (\cite{brown}). As it turns out, the
final sample only contains stars with $\sigma_\pi/\pi < 0.08$,
strongly suggesting that the stars whose true parallax are
systematically smaller than the observed parallax are not
preferentially included in our sample.

To study the possible biases which remain in our subdwarf sample, we
have constructed a Monte Carlo simulation to generate synthetic data
whose properties are known, and compared to `observed' properties
which are calculated in the Monte Carlo.  This is similar in spirit to
the bias studies of \cite{gratton2} and \cite{pont}.  We have
attempted to construct a subdwarf data set whose properties and 
selection biases closely match those in our actual data set. In
particular, our subdwarf sample consists of stars with $\feh < -1.0$,
${\rm M_V} > 5.5$, and $\sigma_\pi/\pi < 0.1$ and these facts are
incorporated in the Monte Carlo.
The Monte Carlo was constructed in the following steps:
\begin{enumerate}

\item An intrinsic \feh\ value (below $\feh = -1.0$) was drawn from one
of two probability functions.  The first function is that given by the
observed \feh\ distribution in the \cite{clla} study
\beq
P (\feh) = 47.13 + 14.35\feh.
\label{fehcarney}
\eeq
This function was chosen as many of the metal-poor stars in the
Hipparcos input catalog are in the \cite{clla} \feh\ catalogue.  The
second \feh\ distribution we considered was that given by \cite{pont}
\beq
P (\feh) = 1.4\exp(\feh + 3) - 1
\label{fehpont}
\eeq
and represents their approximation to the observed \feh\ distribution
in their sample.

\item An observed \feh\ value  ${\rm [Fe/H]}_o$
was determined from the intrinsic \feh\ value by adding a random value
which was taken from gaussian distribution with $\sigma_{\rm [Fe/H]} =
0.10,\; 0.15$ and $0.20$.

\item The distance $d$ was determined assuming a sphere of uniform
density and the true parallax was determined, $\pi_t = 1/d$.

\item The absolute magnitude ($\rm M_V$) was determined assuming an
Salpeter initial mass function  ($\Phi(m) \propto m^{-2.35}$, with
upper and lower mass limits taken to be $m = 0.9$ and $0.4\,\msun$)
and a mass luminosity relation taken from our standard isochrones:
\beqa
{\rm M_V} &=& 13.81 - 12.11m,\;\;\; {\rm for}\; \feh < -1.5\;
{\rm and} \nonumber\\
{\rm M_V} &=& 15.14 - 13.02m\;\;\; {\rm for}\; -1.0 \le \feh \le -1.5.
\eeqa

\item The true apparent magnitude $V_t$ was calculated from the absolute
magnitude and true parallax $V_t = {\rm M_V} - 5.0\log(\pi_t) - 5.0$.

\item The observed magnitude $V_o$ was calculated from the apparent
magnitude assuming gaussian errors with $\sigma_V = 0.02$

\item To reproduce the Hipparcos catalogue completeness
characteristics, we assumed the catalogue was complete up to $V_o = 9$
and increasingly incomplete fainter than this, with a probability for
inclusion of 
\beq
P(V) = 10^{-\tau({V_o} - 9)}
\eeq
where $\tau$ was chosen to be $0.6$ which is valid for the entire
Hipparcos catalogue (\cite{pont}), or $\tau = 0.15$ which is valid for
the metal-poor stars in the \cite{clla} \feh\ catalogue.

\item The observed parallax ($\pi_o$) was computed from the true
parallax assuming gaussian errors with $\sigma_\pi$.  The value of
$\sigma_\pi$ was taken from a fit to the parallax errors in the
Hipparcos catalogue
\beq 
\sigma_\pi = -3.96 + 1.893V_o - 0.26465*V_o^2 + 0.013107V_o^3
\label{eqsigpi}
\eeq
in units of mas.
The scatter of the parallax errors about the mean value given by
equation (\ref{eqsigpi}) was taken into account by adding 
a random value to $\sigma_\pi$ which was taken from gaussian
distribution with a $\sigma$ varying from  0.54 to 1.14 mas between 
$V_o = 5$ to 12.  This derived parallax error will be refered to as 
$\sigma_{\pi_o}$.
This procedure accurately reproduces the observed
parallax errors as a function of apparent magnitude found in Hipparcos
catalogue.  

\item The observed absolute magnitude ${\rm M_V}_o$ was calculated 
\beq
{\rm M_V}_o = V_o + 5.0\log(\pi_o) + 5.0.
\eeq

\item The simulated data are accepted if 
(a) $\sigma_{\pi_o}/ \pi_o < 0.10$, (b) ${\rm M_V}_o > 5.5$ and (c)
${\rm [Fe/H]}_o$ was below some value.  The
\feh\ cuttoff was allowed to vary, so that some runs required 
 ${\rm [Fe/H]}_o < -1.0$ (valid for our sample), and others required 
${\rm [Fe/H]}_o < -1.8$, the cutoff used by \cite{pont}
\end{enumerate}

A typical simulation contained $10^7$ simulated input stars, of which
$\sim 10^4$ were accepted.  For the data which was accepted, the mean
absolute magnitude and \feh\ biases were calculated.  Both the
weighted mean, and unweighted mean bias was calculated.  As we use
weighted fits in the subdwarf fitting analysis, it is the weighted
mean bias which is appropriate for our sample.  However, \cite{pont}
determined an unweighted mean bias, so this quantity was calculated as
well in order to compare our results to \cite{pont}.  The results are
summarized in Table \ref{tab7} for the various cases given above.  In
all cases, we found that the weighted mean absolute magnitude bias was
small. The largest (in absolute value) absolute magnitude weighted
bias was $-0.006$ mag.  This translates into an age reduction of less
than 0.1 Gyr.  Given the small value of this correction, we have
elected not applied it to our subdwarf fitting.  Our results are in
good agreement with those of \cite{gratton2}.  The unweighted mean
absolute magnitude bias is typically a factor of 10 larger than the
weighted mean, but is still relatively small (maximum absolute value
of $-0.034$ mag).  In no case did we find a positive absolute magnitude
bias (as was found by \cite{pont}).

The weighted mean \feh\ bias could be as large as $+0.12$ dex for
stars selected to have $\feh < -1.8$. However, we did not use any
stars with $\feh < -1.8$ in our main sequence fitting
analysis. Considering the samples with have a metallicity cut at 
$\feh < -1.0$, the weighted mean \feh\ bias is likely to lie in the
range $+0.01$ to  $+0.08$ dex.  We believe case B best represents the
true subdwarf distribution; it has a weighted mean \feh\ bias of
$+0.03$ dex, which (considering the results presented in \S
\ref{sectn6752} to  \ref{sectm13}) could result in an absolute
magnitude bias correction up to $+0.015$ mag. Given the small value of
this correction, and the fact that it acts in the opposite sense to
the absolute magnitude bias correction determined above, we have
elected not to apply it to our main sequence fitting results.  Our
main sequence fitting results allow for up to a $+0.20$ dex systematic
error in the subdwarf metallicity scale.

\clearpage

\clearpage

\begin{deluxetable}{llllll}
\tablecaption{Astrometric Distances \label{tab1}}
\tablewidth{0pt}
\tablehead{
\colhead{Cluster} & 
\colhead{\feh}   & 
\colhead{${\rm (m - M)_O}$}  &
\colhead{V(HB)}   & 
\colhead{$\rm M_V(HB)$}  &
\colhead{\feh\ reference} 
} 

\startdata
M5   &   $-1.17$ & 14.44 & $15.05$  &   $0.51 \pm  0.41$  
&\cite{snedenm5}\nl
M4   &   $-1.33$ & 11.18 & $13.37$  &   $0.67 \pm  0.23$ &\cite{zinnw}  
\nl
M22  &   $-1.75$ & 12.17 & $14.10$  &   $0.58 \pm  0.19$ &\cite{zinnw}  
\nl
M3\tablenotemark{a}   				  

     &   $-1.47$ & 14.91 & $15.63$  &   $0.69 \pm  0.59$ &  
\cite{snedenm3}\nl
M13\tablenotemark{b} 

     &   $-1.58$ & 14.06 & $14.83$  &   $0.71 \pm  0.23$ &  
\cite{kraft}\nl
M92  &   $-2.25$ & 14.76 & $15.13$  &   $0.31 \pm  0.32$ &  
\cite{snedenm92}\nl
\enddata
\tablenotetext{a}{V(HB) from  \cite{buon}. Adopted reddening of 0.01
from \cite{zinn}. }
\tablenotetext{b}{V(HB) from \cite{bcf}}.
\end{deluxetable}

\begin{deluxetable}{lrcclrll}
\tablecaption{Calibrating Subdwarfs\label{tab2}}
\tablewidth{0pt}
\tablehead{
\colhead{Name} & 
\colhead{${\rm V_O}$}   & 
\colhead{${\rm (B-V)_O}$}  &
\colhead{\ebv}   & 
\colhead{\feh}  &
\colhead{$\pi$ (mas)}  &
\colhead{$\sigma_\pi/\pi$} & 
\colhead{${\rm M_V}$}
} 
\startdata
HD 193901  &  8.65 & 0.56 & 0.00 & $-1.07$&  22.88 & 0.054 & $5.45\pm 0.117$\nl
HD 145417  &  7.53 & 0.82 & 0.00 & $-1.15$&  72.75 & 0.011 & $6.84\pm 0.024$\nl
HD 103095  &  6.43 & 0.75 & 0.00 & $-1.22$& 109.21 & 0.007 & $6.62\pm 0.015$\nl
HD 120559  &  7.97 & 0.66 & 0.00 & $-1.23$&  40.02 & 0.025 & $5.98\pm 0.054$\nl
HD 126681  &  9.28 & 0.61 & 0.00 & $-1.45$&  19.16 & 0.075 & $5.69\pm 0.163$\nl
BD+59 2407 & 10.20 & 0.58 & 0.05 & $-1.75$&  15.20 & 0.080 & $6.11\pm 0.174$\nl
HD  25329  &  8.51 & 0.87 & 0.00 & $-1.76$&  54.14 & 0.020 & $7.18\pm 0.043$\nl
CPD-80 349 & 10.05 & 0.54 & 0.02 & $-2.26$&  16.46 & 0.060 & $6.13\pm 0.130$\nl
\enddata
\end{deluxetable}

\begin{deluxetable}{lllc}
\tablecaption{\mvrr\ Calibration\label{tab3}}
\tablewidth{0pt}
\tablehead{
&&&
\colhead{\mvrr\ at}    \nl
\colhead{Method} & 
\colhead{\feh}   & 
\colhead{\mvrr}  &
\colhead{$\feh = -1.9$}    
} 

\startdata
Astrometric & $-1.59$ & $0.59\pm 0.11$ & $0.52\pm 0.11$\nl
White dwarf fitting to N6752 & $-1.51$ & $0.45\pm 0.14$ & $0.36\pm  
0.14$ \nl
Subdwarf fitting to N6752 & $-1.51$ & $0.30\pm 0.15$ & $0.21\pm 0.15$  
\nl
Subdwarf fitting to M5 & $-1.17$ & $0.54\pm 0.09$ & $0.37\pm 0.09$\nl
Subdwarf fitting to M13 & $-1.58$ & $0.36\pm 0.14$ & $0.29\pm 0.14$\nl
LMC RR Lyr & $-1.90$ &$0.44\pm 0.14$ &$0.44\pm  
0.14$\nl
Theoretical models  & $-2.20$ & $0.36\pm 0.10$ & $0.43\pm 0.10$\nl
\enddata
\end{deluxetable}

\begin{deluxetable}{lll}
\tablecaption{Monte Carlo Input Parameters\label{tab4}}
\tablewidth{0pt}
\tablehead{
\colhead{Parameter}&
\colhead{Distribution}&
\colhead{Comment}
}
\startdata
mixing length & $1.85\pm 0.25$ (stat.) & fits GC observations\nl
helium diffusion coefficients & 0.3 -- 1.2 (syst.) & possible
systematic error dominate\nl
high temperature opacities & $1\pm 0.01$ (stat.) & comparison of
OPAL\nl
~~ & & \& LAOL opacities\nl
low temperature opacities & $0.7 - 1.3$ (syst.) & comparison of  
different tables\nl
primordial $^4$He abundance & $0.22 - 0.25$ (syst.)&possible
systematic error dominate\nl
oxygen abundance, [O/Fe] & $+0.55\pm 0.05$ (stat.) & mean from  
\cite{nissen}\nl
& $\pm 0.20$(syst.)\nl
surface boundary condition & \multicolumn{2}{l}{grey or \cite{kris}}\nl
colour table & \multicolumn{2}{l}{\cite{ryi} or \cite{kurcol}}
\nl
Nuclear Reaction Rates:\nl
$p + p \longrightarrow {^2\rm H} + e^+ + \nu_e$ & $1\pm 0.002$ (stat.)
& see \cite{first}\nl
& $^{+0.0014}_{-0.0009}\,\,^{+0.02}_{-0.012}$ (syst.) \nl
${\rm {^3He} + {^3He} \longrightarrow {^4He}} + 2p$ & $1\pm 0.06$  
(stat.)&
\cite{bahcpin} \nl
${\rm {^3He} + {^4He} \longrightarrow {^7Be}} + \gamma$ & $1\pm 0.032$  
(stat.)
& \cite{bahcpin}\nl
${\rm {^{12}C} + p \longrightarrow {^{13}N}} + \gamma$ & $1\pm 0.15$  
(stat.) 

& \cite{bahcall}\nl
${\rm {^{13}C} + p \longrightarrow {^{14}N}} + \gamma$ & $1\pm 0.15$  
(stat.) 

& \cite{bahcall}\nl
${\rm {^{14}N} + p \longrightarrow {^{15}O}} + \gamma$ & $1\pm 0.12$  
(stat.)
& \cite{bahcall}\nl
${\rm {^{16}O} + p \longrightarrow {^{17}F}} + \gamma$ & $1\pm 0.16$  
(stat.)
& \cite{bahcall}\nl

\enddata
\end{deluxetable}

\begin{deluxetable}{llllll}
\tablecaption{Fit coefficients for age as a function of \mvbto\ and
\mvto\ \label{tab5}}
\tablewidth{0pt}
\tablehead{
&
\colhead{$\beta_1$} & 
\colhead{$\beta_2$} & 
\colhead{$\beta_3$} & 
\colhead{$\beta_4$} & 
\colhead{$\beta_5$} 
} 

\startdata
\mvbto\ median & $-0.824$  & $0.418$  & $-0.248$ & $-0.033$ &  
$-0.014$\nl
\mvbto\ $+1\,\sigma$& $-0.775$& $0.418$& $-0.221$& $-0.030$&  
$-0.017$\nl
\mvbto\ $-1\,\sigma$& $-0.857$& $0.413$& $-0.266$& $-0.037$&
$-0.014$\nl
\mvto\ median & $-1.305$& $0.515$& $-0.396$& $-0.018$& $0.049$\nl
\mvto\ $+1\,\sigma$& $-1.322$& $0.524$& $-0.428$& $-0.024$& $0.052$\nl
\mvto\ $-1\,\sigma$& $-1.285$& $0.505$& $-0.361$& $-0.011$& $0.044$\nl
\enddata
\end{deluxetable}

\begin{deluxetable}{llllr}
\tablecaption{GC distances and Ages \label{tab6}}
\tablewidth{0pt}
\tablehead{
\colhead{Name} & 
\colhead{\feh} & 
\colhead{\ebv} & 
\colhead{${\rm (m-M)_V}$} & 
\colhead{Age (Gyr)}  
} 
\startdata
NGC 6752 & $-1.51\pm 0.08$  & $0.04\pm 0.01$ & $13.25\pm 0.07$ &
          $11.2\pm 1.3$\nl 
M5 & $-1.17\pm 0.08$ & $0.03\pm 0.01$ & $14.51\pm 0.09$ & $8.9\pm 1.1$\nl
M13 &$-1.58\pm 0.08$ & $0.02\pm 0.01$ & $14.47\pm 0.09$ & $10.9\pm 1.4$\nl
\enddata
\end{deluxetable}

\begin{deluxetable}{llcclrcrc}
\tablecaption{Monte Carlo Subdwarf Bias Results  \label{tab7}}
\tablewidth{0pt}
\tablehead{
&&&&&
\multicolumn{2}{c}{weighted mean bias} & 
\multicolumn{2}{c}{unweighted mean bias}\\
& 
\colhead{\feh} &  & &
\colhead{\feh} &
\colhead{${\rm M_V}$} &   
\colhead{\feh} &   
\colhead{${\rm M_V}$} &   
\colhead{\feh} \\
\colhead{Case} & 
\colhead{distribution} & 
\colhead{$\sigma_{\rm [Fe/H]}$} & 
\colhead{$\tau$} & 
\colhead{cut} &
\colhead{(mag)} &   
\colhead{(dex)} &   
\colhead{(mag)} &   
\colhead{(dex)} 
} 
\startdata
A & \cite{clla} & 0.10 & 0.15 & $< -1.0$ & $-0.003$ & $+0.011$ &
$-0.034$ & $+0.004$\nl
B & \cite{clla} & 0.15 & 0.15 & $< -1.0$ & $-0.004$ & $+0.033$ &
$-0.034$ & $+0.009$\nl
C & \cite{clla} & 0.20 & 0.15 & $< -1.0$ & $-0.005$ & $+0.057$ &
$-0.034$ & $+0.016$\nl
D & \cite{clla} & 0.10 & 0.60 & $< -1.0$ & $-0.003$ & $+0.022$ &
$-0.025$ & $+0.003$\nl
E & \cite{clla} & 0.20 & 0.60 & $< -1.0$ & $-0.004$ & $+0.064$ &
$-0.026$ & $+0.013$\nl
F & \cite{pont} & 0.15 & 0.15 & $< -1.0$ & $-0.004$ & $+0.033$ &
$-0.034$ & $+0.009$\nl
G & \cite{pont} & 0.20 & 0.15 & $< -1.0$ & $-0.006$ & $+0.074$ &
$-0.031$ & $+0.025$\nl
H & \cite{pont} & 0.20 & 0.60 & $< -1.0$ & $-0.005$ & $+0.077$ &
$-0.024$ & $+0.019$\nl
I & \cite{pont} & 0.15 & 0.60 & $< -1.8$ & $-0.004$ & $+0.076$ &
$-0.028$ & $+0.047$\nl
J & \cite{pont} & 0.20 & 0.60 & $< -1.8$ & $-0.005$ & $+0.125$ &
$-0.028$ & $+0.077$\nl
\enddata
\end{deluxetable}

\clearpage

\clearpage

\end{document}